\documentclass[letterpaper,11pt]{article}

\usepackage[numbers]{natbib}

\usepackage{tabularx} % extra features for tabular environment
\usepackage{float}
\usepackage{caption}
\usepackage{subcaption}
\usepackage{amsmath}  % improve math presentation
\usepackage{graphicx} % takes care of graphic including machinery
\usepackage{multirow}
\usepackage{xcolor}
\usepackage[normalem]{ulem}
\usepackage[margin=1in,letterpaper]{geometry} % decreases margins
\usepackage[final]{hyperref} % adds hyper links inside the generated pdf file
\hypersetup{
	colorlinks=true,       % false: boxed links; true: colored links
	linkcolor=blue,        % color of internal links
	citecolor=blue,        % color of links to bibliography
filecolor=magenta,     % color of file links
urlcolor=blue         
}

\begin{document}

\title{DISTRIBUTED ACOUSTIC SENSING \\\textbf{Distributed Acoustic Sensing for Environmental Monitoring, and Newtonian Noise Mitigation: Comparable Sensitivity to
		Seismometers}}

\author{
	Reinhardt Rading\textsuperscript{1}\thanks{Corresponding author: \textcolor{blue}{\uline{r.r.rading@ieee.org}}}, \
	Fracensca Badaracco\textsuperscript{2},
	Spiridon Beis\textsuperscript{3},
	Katharina Sophie Isleif\textsuperscript{1},
	Paul Ophardt\textsuperscript{1},\\
	Wanda Vossius\textsuperscript{1},
	and the WAVE Collaboration\textsuperscript{1,4,5,6,7}
}

\maketitle

%------------------------------------------ Author Affiliations ----------------------------------------------%

\begin{center}

	\small
	\textsuperscript{1} Helmut Schmidt University, Hamburg, Germany \\
	\textsuperscript{2} Università degli Studi di Genova, Genoa, Italy\\
	\textsuperscript{3} RWTH Aachen University, Aachen, Germany \\
	\textsuperscript{4} University of Hamburg, Hamburg, Germany \\
	\textsuperscript{5} Deutsches Elektronen-Synchrotron, Hamburg, Germany \\
	\textsuperscript{6} European XFEL, Schenefeld, Germany \\
	\textsuperscript{7} GFZ Potsdam, Potsdam, Germany \\
\end{center}

\vspace{0.5cm}

\begin{abstract}
\noindent Newtonian noise limits the low-frequency sensitivity of ground-based gravitational wave detectors. While seismometers and geophones are commonly employed to monitor ground motion for Newtonian noise cancellation, their limited spatial coverage and high deployment costs hinder scalability. In this study, we demonstrate that distributed acoustic sensing offers a viable and scalable alternative, providing performance comparable to that of conventional seismic instruments. Using data from acoustic sensing and colocated seismometers during both natural and controlled events, we observe a strong correlation (exceeding 0.8) between the two sensor types in the 3–20 Hz frequency band relevant for Newtonian noise. Moreover, when distributed acoustic sensing data are used to predict geophone signals, the correlation remains high (above 0.7), indicating that distributed acoustic sensing accurately captures both the spatial and spectral features of ground motion. As a case study, we apply distributed acoustic sensing data to cancel noise recorded by the vertical component of a seismometer and compare the results with those obtained using geophone data for the same task. Both distributed acoustic sensing- and geophone-based cancellations yield a residual noise factor of 0.11 at 20 Hz. These findings confirm the feasibility of using distributed acoustic sensing for Newtonian noise mitigation and highlight its potential, in combination with traditional seismic sensors, to improve environmental monitoring and noise suppression in current and next-generation gravitational wave observatories.
\end{abstract}

\section{Introduction}
Distributed Acoustic Sensing (DAS) is an emerging technology that transforms standard optical
fibers into dense arrays of virtual sensors. Using coherent backscattering of light within the
fiber, DAS can provide thousands of strain measurements per kilometer with meter-scale spatial
resolution \cite{b1}. This makes DAS suited for monitoring ground motion over large coverage areas
and with fine spatial detail. Despite its promise, DAS has often been considered inferior to
seismometers for ground motion sensing due to concerns about its sensitivity, calibration accuracy, and response to different wave types \cite{b2a} \cite{b2b}. Previous comparisons have shown that DAS can
capture strong seismic events \cite{b3}, but detailed, validation of its performance in Newtonian noise \cite{b2a}\cite{b4}
cancellation remains limited.
This study addresses that gap. We directly compare the performance of DAS and colocated
broadband seismometers and geophones during both natural and controlled seismic events. By analyzing
data from a regional earthquake and a vibrotruck experiment, we show that DAS can match
seismometer readings in both amplitude and phase, and can accurately predict geophone
output using Wiener filters and DAS measurements. These findings support the hypothesis that DAS is not
only suitable but also competitive with seismometers for Newtonian noise cancellation and environmental
monitoring, especially in large-scale detector deployments like the Einstein Telescope.

\section{DAS: Principle of Operation}
\noindent DAS is an interferometric sensing technique that transforms a standard optical fiber into a dense
array of virtual sensors capable of measuring dynamic strain or strain rate along its length. DAS operates based on Rayleigh backscattering, which arises from microscopic fluctuations in the refractive index of the optical fiber. These fluctuations originate during the fiber manufacturing process—particularly during preform fabrication and the rapid cooling that follows fiber drawing—leading to non-uniformities in the core material\cite{b15a}. Since these fluctuations occur on scales much smaller than the optical wavelength, illuminating the fiber with light of a wavelength significantly larger than the scattering centres results in Rayleigh scattering.

\vspace{0.1cm}

\noindent When coherent
laser pulses are injected into the fiber, a small fraction of the light is scattered in all directions.
The part of this light that is scattered back toward the source (backscattered light) carries
information about the local optical path length and, consequently, any mechanical perturbations
experienced by the fiber. The most commonly employed interrogation method is Optical Time Domain Reflectometry (OTDR) \cite{b16a}. In this scheme, a series of narrow, coherent optical pulses is launched into one end of the fiber. As the pulses propagate, Rayleigh-scattered light returns continuously to the source, where it is detected and analyzed. Because the travel time of the backscattered signal is proportional to the distance from the launch point, the spatial location of scattering events can be inferred through precise timing, enabling distributed sensing along the entire fiber.

\vspace{0.1cm}

\begin{figure}[htbp]
	\centering\includegraphics[width=15cm]{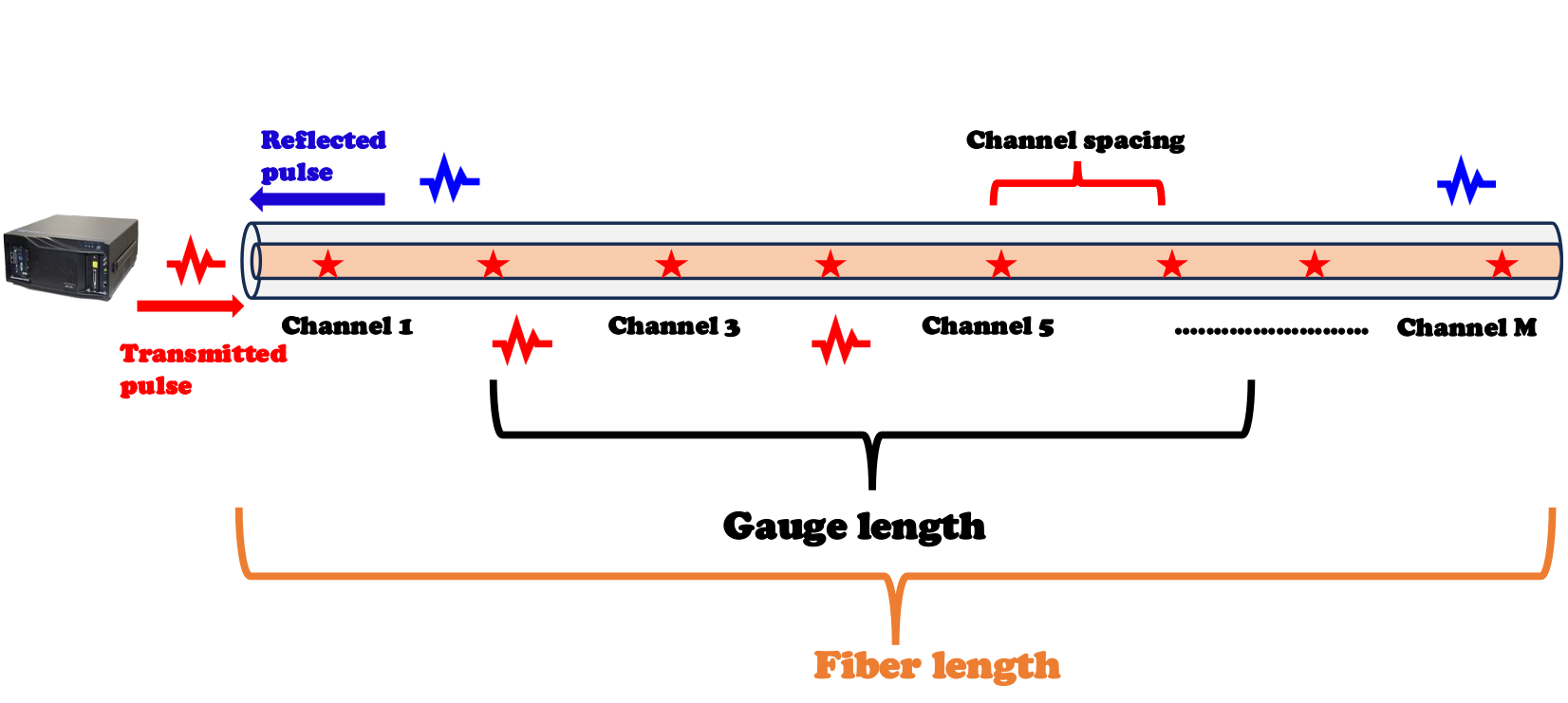}
	\caption{\textit{Schematic representation of a distributed optical fiber sensor. The fiber can be defined as consisting of discrete virtual sensing channels (points) with a defined channel spacing (spatial resolution) and gauge length. Local phase changes are estimated by differentiating the phase at each fiber channel with respect to the preceding one, using the first channel as a reference. Repeating this process over time enables the detection and localization of mechanical perturbations along the fiber.}}
	\label{fig1}
\end{figure}

\noindent To better understand DAS, it is useful to define some key terms. First, the concept of a sensor array is fundamental: in the case of distributed sensors (interchangeably used as channels), the system functions as an array of sensing points rather than a single sensor (channel), as illustrated in  figure  \ref{fig1}. The spatial resolution (channel spacing) refers to the smallest distance along the sensor array at which two distinct measurand values can be resolved from the backscattered light \cite{b14a}. The instrument configuration includes an interrogation frequency and a sampling frequency at the receiver, which together define the spatial sampling resolution. This sampling resolution may differ significantly from the true spatial resolution of the sensor. Finally, the gauge length is the effective spatial resolution over which the system takes the measurement and is explained in more detail in the following section.

\subsection{Sensitivity to Strain and Phase Measurement}
When the optical fiber is subject to local deformation—for example, due to seismic activity, acoustic waves, or structural stress—these mechanical disturbances alter the local refractive index and geometry of the fiber \cite{b1}. Such perturbations introduce a differential optical phase shift between portions of the pulse scattered before and after passing through the disturbed zone.
\noindent DAS systems typically extract this phase shift by comparing the backscattered signals from two temporally separated probe pulses using an interferometric setup \cite{b1}. The two pulses are delayed in time, often via an optical delay loop, which defines an effective gauge length \begin{math}
	L_{g}
\end{math}. This gauge length is given by \cite{b1}:
\begin{equation}
	L_{g}=\frac{V_{f} \Delta t}{2},
\end{equation} where \begin{math}
	V_{f}
\end{math} is the speed of light in the fiber and \begin{math}
	\Delta t
\end{math} is the temporal separation between the pulses. The phase change \begin{math}
	\Delta \phi
\end{math} measured between the two pulses is proportional to the change in optical path length over the gauge length and hence to the average axial strain \begin{math}
	\varepsilon 
\end{math} or strain rate \begin{math}
	\frac{d\varepsilon}{dt}
\end{math} along that segment of fiber.

\subsection{What DAS Measures: Strain (strain rate) along the Fiber}

It is important to note that DAS does not provide true point measurements. Instead, the measured strain (or strain rate) is spatially averaged over the gauge length. As a result, the change in displacement \begin{math}
	\Delta 	\boldsymbol{u}
\end{math} measured by DAS at a given point  \begin{math}
	x
\end{math} between time steps \begin{math}
	t
\end{math}  and \begin{math}
t+dt
\end{math} is \cite{b9a}-\cite{b11a}:
\begin{equation}
\Delta 	\boldsymbol{u}= \left[ 	\boldsymbol{u}(x+\frac{L_{g}}{2},t+dt)-	\boldsymbol{u}(x-\frac{L_{g}}{2},t+dt)  \right]-\left[ 	\boldsymbol{u}(x+\frac{L_{g}}{2},t)-	\boldsymbol{u}(x-\frac{L_{g}}{2},t)  \right].
\label{eq:phase_difference}
\end{equation}  
Dividing Eq.~\eqref{eq:phase_difference} by \begin{math}
	L_{g}
\end{math} and \begin{math}
dt
\end{math}, gives the strain-rate over the gauge length.  The choice of gauge length is thus critical: a longer gauge length increases the signal-to-noise ratio (SNR) by suppressing  noise components but at the cost of spatial resolution. Conversely, a shorter gauge length enhances spatial resolution but may suffer from lower SNR. In this sense, the gauge length acts analogously to a moving average filter, smoothing the recorded signal along the fiber. 
\vspace{0.2cm}

\begin{figure}[htbp]
	\centering\includegraphics[width=8cm]{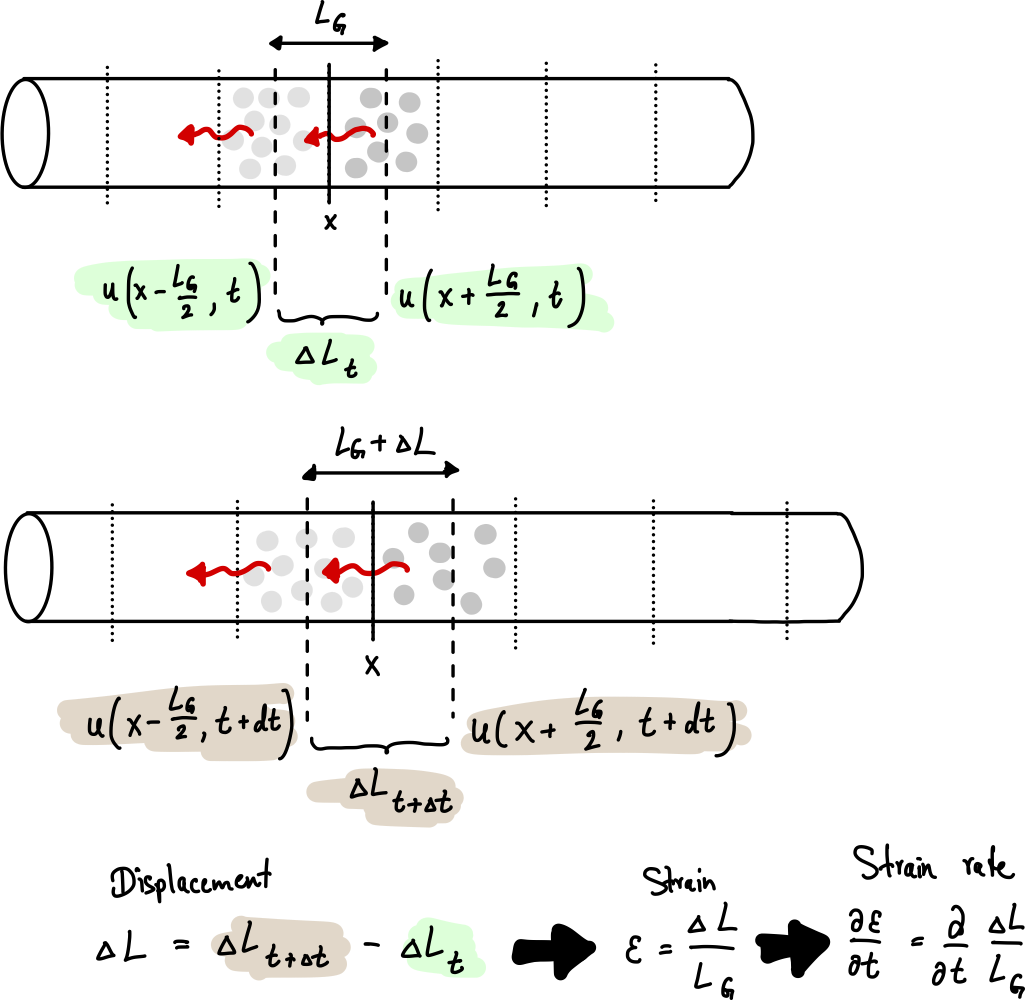}
	\caption{\textit{Illustration of the displacement measured by the DAS system at a given position $x$, showing the change between two time steps, $t+\Delta t$.
		}}
	\label{fig23}
\end{figure}

\noindent The strain rate can be converted to strain by integrating over time.
This strain is a one-dimensional projection of the full strain tensor $\varepsilon$.
\noindent The strain tensor is derived from the displacement field \begin{math}
	\boldsymbol{u}_{i}(x,t) 
\end{math} as:
\begin{equation}
	\varepsilon_{ij} = \frac{1}{2} \left( \frac{d	\boldsymbol{u}_{i}}{dx_{j}} + \frac{d	\boldsymbol{u}_{j}}{dx_{i}} \right), 
\end{equation} where \begin{math}
		\boldsymbol{u}_{i}
\end{math} and \begin{math}
		\boldsymbol{u}_{j}
\end{math} are displacement components in the \begin{math}
	i
\end{math} and  \begin{math}
	j
\end{math} directions, respectively.
\vspace{0.2cm}

\noindent DAS can detect both P-waves (compressional) and S-waves (shear), since both produce strain components along the fiber axis. The sensitivity to a particular wave type depends on the orientation of the fiber relative to the wave propagation direction. For P-waves, which are the compressional (diagonal) components of the strain tensor, any fiber segment aligned with the direction of propagation will experience axial compressional strain. For S-waves, which induce shear deformation, DAS can measure the shear (off-diagonal) components of the strain tensor if the fiber has a non-zero component in the direction of the displacement gradient \cite{b11b}. Even though DAS measures only one scalar strain component per fiber segment, its sensitivity can capture both wave types depending on the geometry; think of helically wound fiber or complex layouts.

\begin{figure}[htbp]
	\centering\includegraphics[width=13cm]{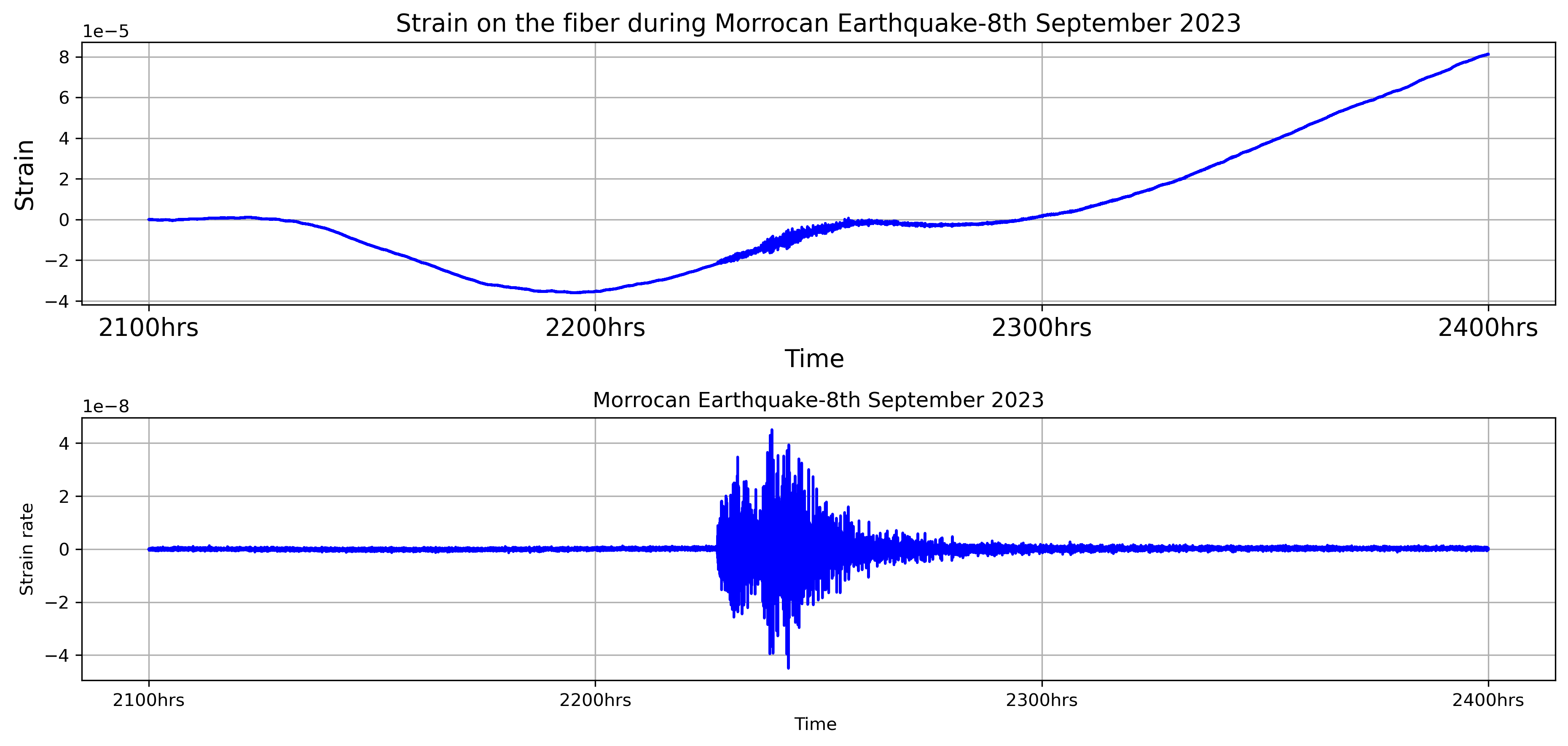}
	\caption{\textit{Distributed Acoustic Sensing (DAS) recording the Morrocan earthquake over 4000 km from Hamburg showing clear arrivals of both compressional (P) and shear (S) waves. The P-wave, arriving first, is characterized by lower amplitude and higher velocity, followed by the higher amplitude S-wave.}}
	\label{fig2}
\end{figure}

\subsubsection{Low-Frequency and Long-Wavelength Sensitivity}
Due to the averaging effect of the gauge length, DAS exhibits reduced sensitivity to low-frequency signals with long spatial wavelengths. However, because DAS enables dense spatial sampling (e.g., 1 m channel spacing over tens of kilometers), low-frequency signals can still be reconstructed by aggregating measurements across multiple channels. In effect, while each individual channel behaves as a spatially averaged strain-rate sensor, the collective response of many closely spaced virtual sensors allows recovery of broader-scale motion.

\section{Equivalence Between DAS and Seismometer Recordings}
For a plane wave propagating in the \begin{math}
	x
\end{math}-direction, displacement \begin{math}
\boldsymbol{u}
\end{math} at point \begin{math}
x
\end{math} at a given time \begin{math}
t
\end{math} is given by: \begin{math}
\boldsymbol{u}(x,t)=\boldsymbol{E}(x)e^{i(kx-\omega t)}
\end{math} and mathematically, the strain rate is defined by:
\begin{equation}
	\frac{d\varepsilon}{dt}=\frac{d}{dt}\left( \frac{d\boldsymbol{u}}{dx}\right),
\end{equation} where  \begin{math}
u
\end{math} is the particle displacement in the cable direction.  In other words, at point \begin{math}
x
\end{math}, DAS is basically performing a finite-difference approximation of the spatial derivative of ground velocity over the gauge length \begin{math}
L_{g}.
\end{math} 

\vspace{0.1cm}
\noindent Most seismometers can measure either displacement or velocity and therefore a relationship between the quantity recorded by DAS and the quantity recorded by a seismometer can be established. It is given by \cite{b10a}-\cite{b11b}:

\begin{equation}
	\frac{d\varepsilon}{dt}=\frac{1}{L_{g}}\int^{x+\frac{L_{g}}{2}}_{x-\frac{L_{g}}{2}}\frac{d}{dl} \frac{d\boldsymbol{u}}{dt} {dl} = \frac{v(x+\frac{L_{g}}{2})-v(x-\frac{L_{g}}{2})}{gl},
\end{equation} where \begin{math}
v
\end{math} is \begin{math}
\frac{d	\boldsymbol{u}}{dt},
\end{math} the velocity as recorded by a seismometer. In simple words, we can convert strain (strain-rate) to seismometer readings (acceleration/velocity/displacement) and vice versa using \cite{b9a}:
\begin{equation}
	\varepsilon = \frac{d\boldsymbol{u}}{dx}=\pm \frac{1}{s} \frac{d\boldsymbol{u}}{dt},
\end{equation} where \begin{math}
s =\frac{\omega}{k}
\end{math} is the apparent local slowness in the cable direction, which can be estimated as described in the following section. 

\subsection{Estimating Apparent Slowness $s$}
Slowness $s$ can be estimated using several approaches. One method involves analyzing the moveout in the time domain from traces recorded along a segment of a DAS cable. Alternatively, it can be determined in the frequency–wavenumber (f–k) domain as the ratio $\frac{\omega}{k}$, where $\omega$ is the wavenumber and $k$ is the angular frequency \cite{b9a}. \noindent However, estimating phase velocities using f-k analysis has several challenges, for example, different seismic phases propagate with different velocities and in different directions, leading to misleading estimation of the value of $s$.  Furthermore, the phase velocity change rapidly over time or space and further assumptions like spatial heterogeneity, such as variations in local velocity structure or changes in the orientation of the fiber, complicates slowness estimation. The techniques used for estimating slowness in this analysis are the semblance method \cite{b6a}-\cite{b8a} and the Multiple Signal Classification (MUSIC) algorithm \cite{b12a}\cite{b13a}. 

\subsubsection{Semblance method}
\begin{figure}[htbp]
	\centering\includegraphics[width=13cm]{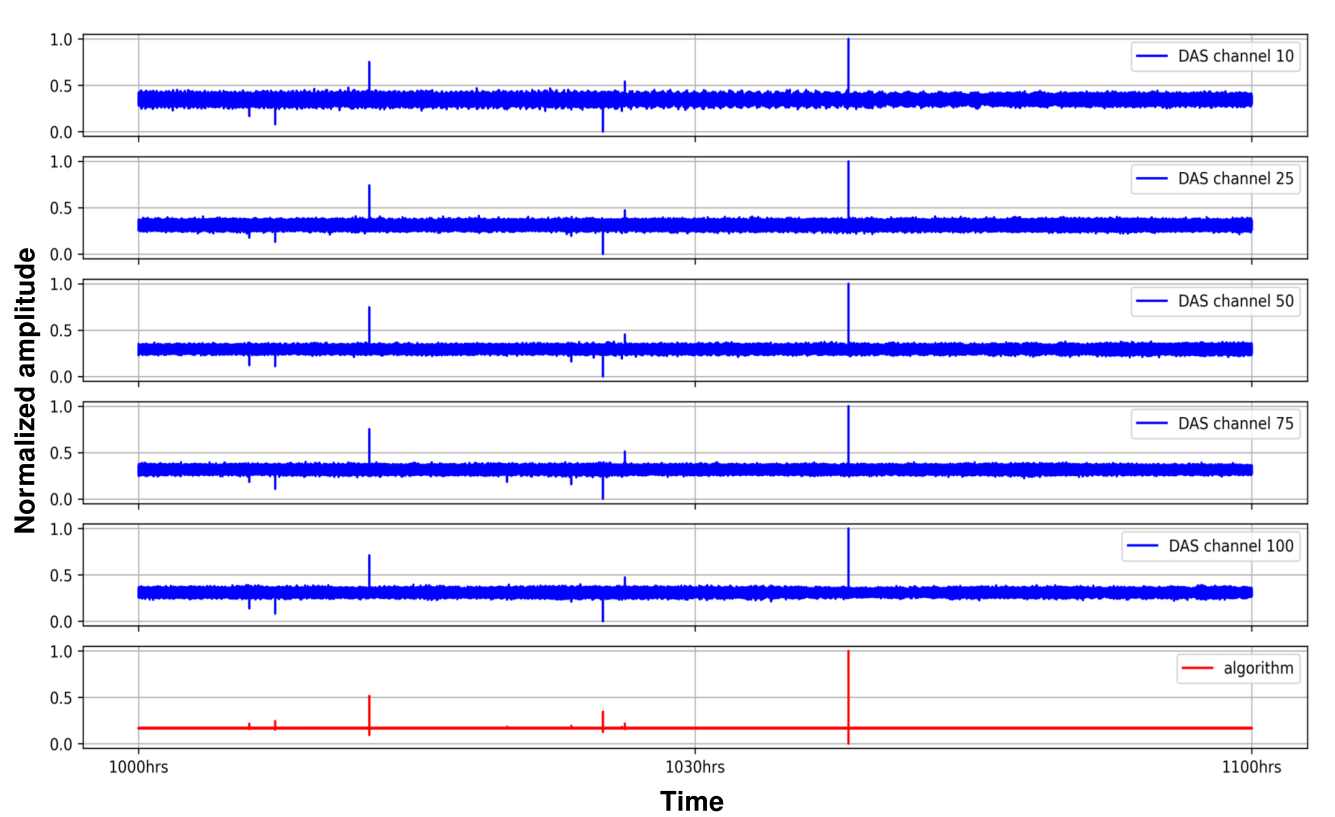}
	\caption{\textit{ Illustration of the semblance algorithm. The algorithm functions as a spatial noise filter based on amplitude scaling, enhancing signal components that are coherent across the sensor array. This filtering is essential for suppressing localized noise that does not exhibit coherence with the rest of the array.}}
	\label{fig3}
\end{figure}

Semblance is an array processing technique used to quantify the coherence of signals recorded by a group of sensors. It measures the degree of similarity among the sensor outputs\cite{b7a}. When the recorded traces are aligned and are similar, the semblance value is high, indicating strong coherence. Conversely, if the traces are misaligned or differ significantly in shape, the semblance value decreases, reflecting lower coherence. Suppose there are  $M$ sensors (see figure \ref{fig1}), the semblance at the $j$-th sample is given by \cite{b8a}:
\begin{equation}
	sem[j]=\frac{[\sum_{i=1}^{M} f(i,j)]^{2}+ [\sum_{i=1}^{M} h(i,j)]^{2}}{M \sum_{i=1}^{M} \{ f(i,j)^{2} + h(i,j)^{2}\} },
	\label{eq7}
\end{equation} where $f(i,j)$ is the $j$-th sample of the $i$-th sensor within a group of $M$ sensors and $h(i,j)$ is the Hilbert transform of $f(i,j)$. The Hilbert transform acts as a phase rotation operator, shifting the phase of the original signal by $90$ degrees at every frequency and time sample. This is essential as it provides more signal processing information related to the sensor array \cite{b6a}\cite{b8a}. 
For $N$ samples, Eq.(\ref{eq7}) can be interpreted as a spatial noise filter based on amplitude scaling. This filtering is important for suppressing local noise that lacks coherence with the rest of the sensor array as shown in figure \ref{fig3}. To ensure accurate signal preservation, it is important to select sensors within the appropriate wavelength range. Specifically, for longer wavelengths, the semblance algorithm must include sensor channels spaced at intervals comparable to or greater than the wavelength to avoid losing relevant signal information.

\noindent To estimate slowness values from the semblance method, Eq.(\ref{eq7}) can be rewritten as (see \ref{intro}):

\begin{equation}
	sem[s_{x},t]=\frac{[\sum_{i=1}^{M} f(t+s_{x}(D_{i}-D_{1}))]^{2}+ [\sum_{i=1}^{M} h(t+s_{x}(D_{i}-D_{1}))]^{2}}{M \sum_{i=1}^{M} \{ f(t+s_{x}(D_{i}-D_{1}))^{2} + h(t+s_{x}(D_{i}-D_{1}))^{2}\} },
	\label{eq8}
\end{equation} where $D_{i}-D_{1}$ is the the separation distance between sensor $i$ and the reference (first) sensor in the array. At each instance in time, a range of slowness values is tested to identify that with the maximum semblance. For each slowness value $s_{x}$, semblance is calculated
and the slowness value with the
highest semblance represents that of the most likely local slowness of a  coherent plane wave at the specific time $t$.

\subsubsection{Multiple Signal Classification (MUSIC) algorithm}
MUSIC is an alternative signal processing technique used to estimate the slowness or arrival direction of an incident wavefield \cite{b12a}. It operates by decomposing the recorded sensor data into orthogonal signal and noise subspaces, enabling  direction-of-arrival (DOA) estimation even in the presence of noise. Consider a set of recordings $Y$ from an array of 
$M$ sensors. The data can be modeled as a linear superposition of signal and noise components:
\begin{equation}
	Y=AX+N.
\end{equation}
The elements of $A$ are known functions of the signal arrival angle
and the array location and $X$ represent the amplitude and phase of the signal.

\noindent From the observed data $Y$, a sample covariance matrix is computed. This covariance matrix is then subjected to Singular Value Decomposition (SVD) or eigenvalue decomposition. The resulting eigenvalues and their corresponding eigenvectors are used to separate the signal and noise subspaces. Eigenvectors associated with eigenvalues exceeding a defined  threshold are considered to span the signal subspace, while the remaining eigenvectors span the noise subspace N. The MUSIC algorithm exploits the orthogonality between the signal subspace and the noise subspace to identify the slowness or arrival angle $\theta$ of coherent signals \noindent and its functionality is given by \cite{b13a}:
\begin{equation}
	\delta_{MUS}(\theta)=\frac{1}{V^{*}(\boldsymbol{k})NN^{*}V(\boldsymbol{k})},
\end{equation} where $V(\boldsymbol{k})$ is the steering vector representing the phase delays at each sensor of the array for an incoming wave with wavenumber vector $\boldsymbol{k}$. The wavenumber vector $\boldsymbol{k}$ is given by
\begin{equation}
		\boldsymbol{k}=2\pi s(f) \cdot (sin\theta, cos\theta,0), 
\end{equation} $s(f)$ describes the slowness of the wave at frequency $f$ and backazimuth $\theta$.

\vspace{0.2cm}

\noindent A limitation of the MUSIC algorithm is its reliance on knowledge of the sensor positions. While this requirement is manageable for conventional  geophone and seismometer arrays, it poses a significant challenge for DAS systems, where the exact location of individual sensing points along the fiber is often uncertain. Additionally, the MUSIC algorithm requires prior knowledge of the number of signal sources in order to accurately separate the events into signal and noise subspaces. Despite these constraints, MUSIC can provide accurate results when these  information are known. In contrast, the semblance method offers a more practical and robust approach, particularly for DAS applications. It does not require knowledge of the precise sensor locations and relies solely on the recorded DAS outputs to evaluate coherence among sensors. Due to its simplicity the semblance method has been the used in this study.

\section{Methods}
\subsection{Experimental Setup}
Our experiments were conducted in collaboration with the WAVE initiative\cite{bb2} on the research campus Bahrenfeld/DESY in Hamburg. The focus of this collaboration is the investigation of large-scale seismic sensor networks and the implementation of fiber-optic sensing technologies for environmental and geophysical monitoring. DAS system is achieved using existing dark fiber installed within underground accelerator tunnels, including those of the European XFEL providing over 19000 seismic sensors along the 19 kilometer fiber. 

\vspace{0.2cm}

\noindent These tunnels provide an ideal, low-noise environment- though noisy compared to gravitational wave detector sites- to assess the performance and sensitivity of DAS for long-term seismic and environmental monitoring. This also allows us to capture a wide range of environmental noise sources, including microseisms, anthropogenic activity (such as vehicles and pedestrian traffic), meteorological effects (e.g., wind or thunder), and transient seismic events (e.g., earthquakes). To benchmark DAS performance, we compare its recordings with co-located conventional seismic sensors, including broadband seismometers and geophones, that are also deployed across the campus. These instruments serve as ground truth references for evaluating DAS sensitivity, signal-to-noise ratio, and its capability for seismic event detection and noise cancellation.  
\vspace{0.2cm}

\noindent In 2021, the WAVE team conducted an active-source campaign using a vibrotruck that generated controlled low-frequency vibrations at various locations across the campus. The ground motions induced by this truck were recorded simultaneously by both the DAS system and the traditional sensors, providing a controlled dataset for comparison. A follow-up campaign took place in 2023, in which a series of stone-drop experiments were conducted at multiple sites on campus. These impulsive events created localized, high-frequency transients that were captured by the DAS system and reference sensors. The recorded waveforms from these events are used to further assess DAS performance, particularly in terms of spatial resolution and response to high-frequency content. 

\vspace{0.1cm}

\noindent For data analysis, the DAS was sampled at 1000 Hz to ensure high temporal resolution, whereas the seismometers operated at 200 Hz. To facilitate direct comparison, the DAS data were resampled to 200 Hz after initial preprocessing, which included detrending and filtering. This preprocessing ensures consistency in temporal and spectral resolution across sensor types and is essential for correlation and coherence analyses.

\begin{figure}[htbp]
	\centering\includegraphics[width=13cm]{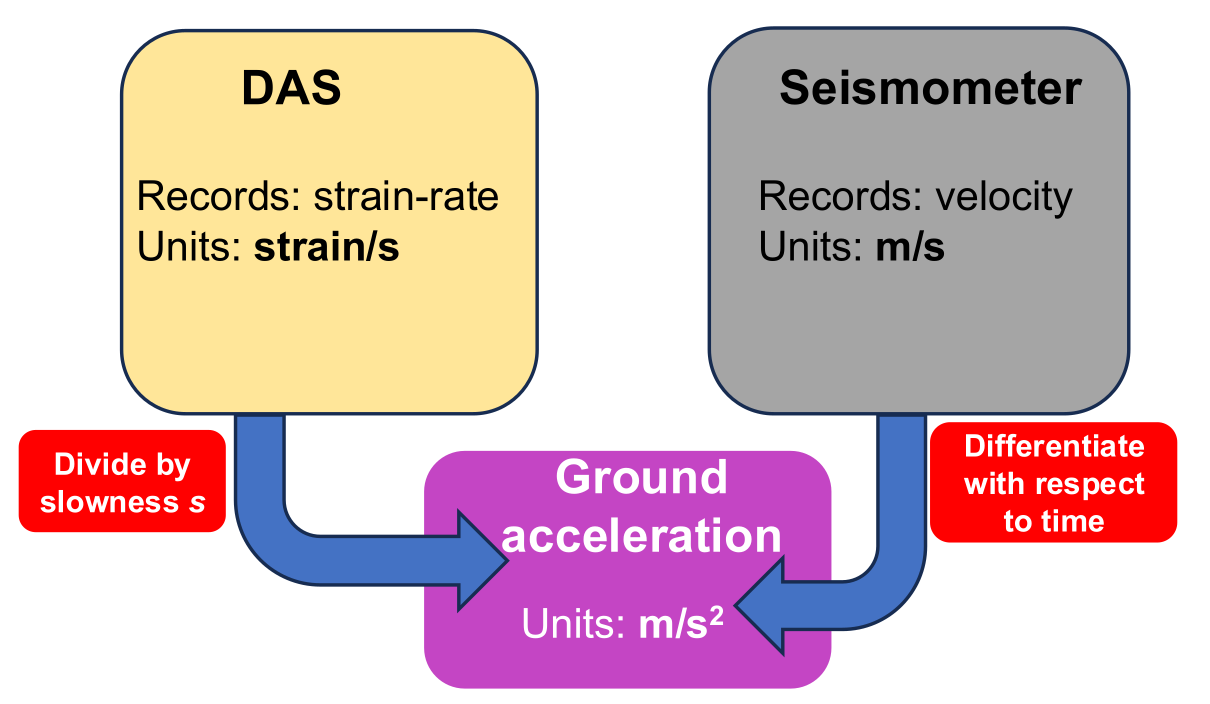}
	\caption{\textit{Illustration of the conversion process used to compare seismometer and DAS measurements in terms of ground acceleration. The DAS output, given as strain rate, is converted to acceleration by dividing by the slowness 
			$s$. The seismometer output, originally in velocity, is converted to acceleration through temporal differentiation.}}
		\label{fig5}
\end{figure}

\vspace{0.1cm}

\noindent The comparison of DAS and seismometer outputs involves converting both recordings to the same physical quantity—in this case, ground acceleration as shown in figure \ref{fig5}. Seismometers  measure ground velocity; differentiating these signals with respect to time yields ground acceleration. Conversely, the DAS system measures strain rate, which is also converted into acceleration. This is achieved by dividing the strain rate by the local slowness, which is estimated dynamically from the DAS data using the semblance (slant-stack) method as described in Eq.(\ref{eq8}). This transformation results in a time series of ground acceleration from the DAS data that is directly comparable to the seismometer-derived acceleration.

\subsection{Reconstructing Geophone Signals from DAS Data}Geophones are widely used in seismology and geotechnical applications to measure ground motion. To demonstrate the potential of DAS for Newtonian noise cancellation and other geophysical applications, we investigate whether DAS can accurately estimate geophone signals using a Wiener filter.
\noindent The Wiener filter, is a optimal linear estimation method used in noise reduction and signal reconstruction. By leveraging the correlation between DAS measurements and geophone data, the Wiener filter aims to minimize the mean square error between the estimated geophone signal and the true signal.

\noindent Mathematically, the Wiener filter can be expressed as a convolution between the input signal $x(t)$- in this case DAS data- and a set of optimal filter coefficients $h(t)$:
\begin{equation}
	\hat{y}(t)=\int_{-\infty}^{\infty} h(\tau) x(t - \tau) \, d\tau,
\end{equation}
where $h(t)$ represents the impulse response of the Wiener filter, and $\hat{y}$ is the estimated signal, which in this case is the geophone signal. The filter coefficients $h(t)$
 are determined by the properties of the input signal and the noise characteristics, aiming to minimize the squared error between $\hat{y}$ and the true geophone signal $y$.
 
 \vspace{0.2cm} 
 
 \noindent In this study, we used data from a 10-channel DAS array, down-sampled to 200 Hz, over a period of 1 hour. The DAS measurements were co-located with a set of geophones, which were also sampled at 400Hz and down-sampled to 200Hz for the same duration. \noindent We implemented the Wiener filter in the frequency domain. The DAS data were first preprocessed to remove low-frequency drifts, normalize the signals, and filter out high-frequency noise components. We assumed stationarity within the data set needed for Wiener filtering.

 \begin{figure}[htbp]
 	\centering
 	\begin{minipage}{0.49\textwidth}
 		\centering
 		\includegraphics[width=\linewidth]{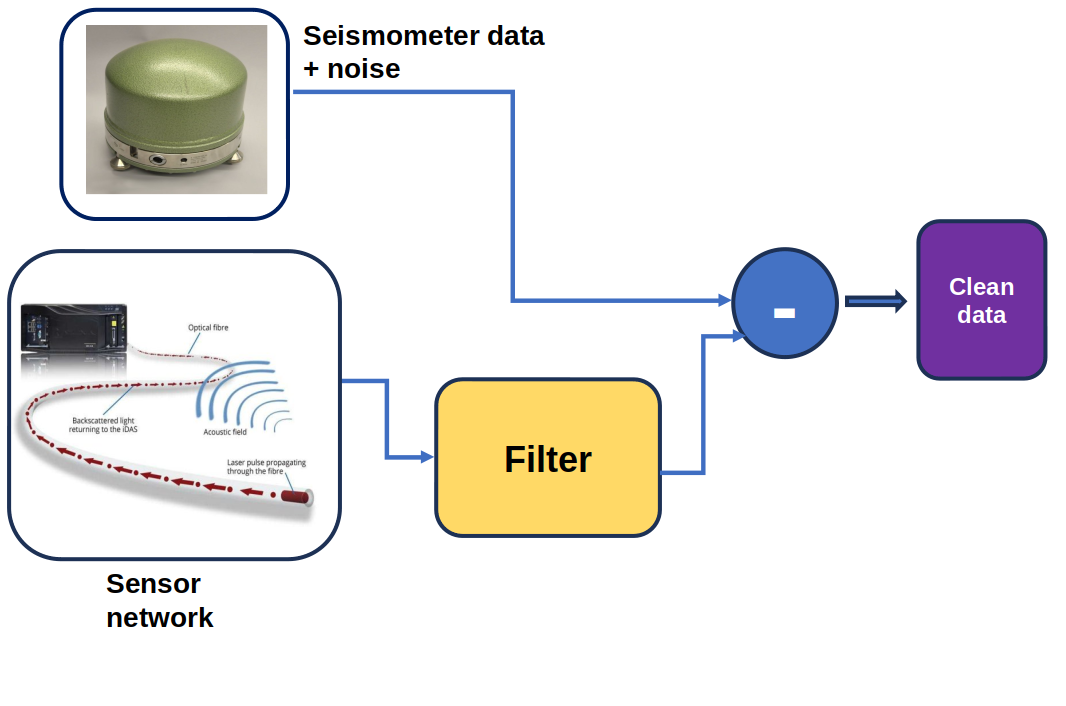}
 	\end{minipage}
 	\hfill
 	\begin{minipage}{0.5\textwidth}
 		\centering
 		\includegraphics[width=\linewidth]{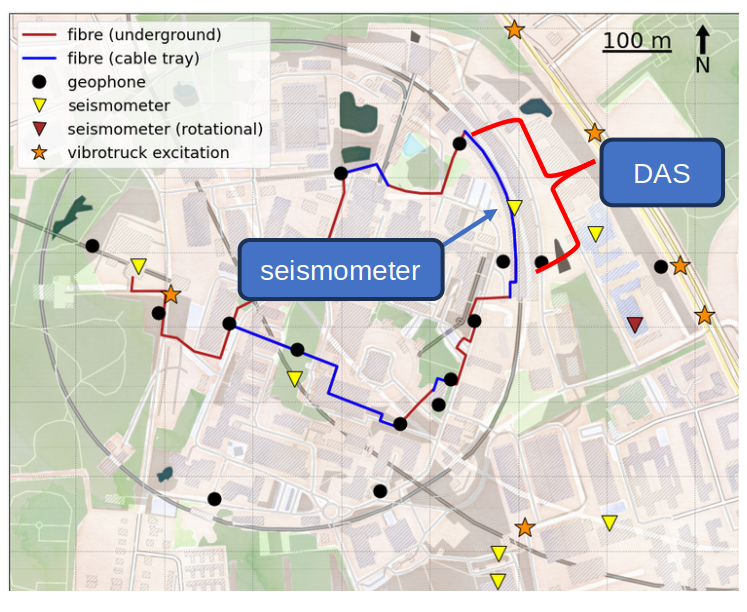}
 	\end{minipage}
 	\caption{\textit{Left:  Block diagram illustrating the process of seismometer noise cancellation$^{\dagger}$. Right:  Map of the DESY campus showing the locations of the seismometer, geophones, and the DAS array, as well as the ground excitation points produced by the vibrotruck. The DAS sensor network highlighted in the figure is used as a case study for cancelling the noise recorded by the vertical component of the seismometer.}}
 	\label{fig68}
 \end{figure}
 
 \subsection{Cancelling Seismometer Noise Using DAS and Geophone Sensor Networks }
 \noindent We extended the concept of Newtonian noise cancellation by using data from both DAS and an array of geophones to suppress noise recorded by a colocated seismometer. We used the DAS array alone to cancel the noise recorded by the seismometer and, separately, used the geophone array for the same purpose. We then compared the performance of both approaches. While geophone-based noise cancellation has been demonstrated previously \cite{b14}, this is the first time, to our knowledge, that DAS has been used to cancel seismic noise in a colocated seismometer. Both the horizontal (East) and vertical components of the seismometer are considered, with particular emphasis on the vertical component, as it is analogous to the test mass motion in gravitational wave detectors. This is done because Newtonian noise tends to correlate more strongly with vertical displacement. However, this relationship holds for surface gravitational wave detectors and not for those located underground.
 
 \vspace{0.1cm}

  \noindent A DAS array was deployed in close proximity to the seismometer, as illustrated in figure \ref{fig68}. In addition, three-component geophones were placed near the seismometer to serve as reference sensors for noise cancellation. The seismometer's vertical channel recorded both the signal of interest and ambient noise. To isolate the signal, we utilized DAS and geophone measurements as reference (witness) channels in the Wiener filter algorithm.
 
 \vspace{0.01cm}
 
\noindent The noise cancellation technique is based on the multichannel Wiener filter and utilizes cross-spectral matrices, expressed as \cite{b14}:
\begin{equation}
	R(\omega) =1 -\frac{\vec{C^{\dagger}}_{DS}(\omega) \cdot \left(\vec{C}_{DD}(\omega)\right)^{-1} \cdot \vec{C}_{DS}(\omega) }{\vec{C}_{SS}(\omega)}
	\label{eq13}, 
\end{equation} where
\begin{math}
	\vec{C}_{DS}
\end{math} is the cross-spectral vector between reference sensors (DAS sensors) and the target sensor (vertical component of the seismometer), \begin{math}
\vec{C}_{DD}
\end{math} is the  cross-spectral matrix of the reference sensors with themselves, \begin{math}
\vec{C^{\dagger}}_{DS}
\end{math} is the Hermitian transpose of \begin{math}
\vec{C}_{DS}
\end{math}. Residual $\sqrt{R(\omega)}$ represents the noise reduction factor that DAS is able to achieve.
\vspace{0.2cm}

\noindent The DAS sensors recorded the strain rate along the horizontal axis of ground motion at a sampling rate of 1000 Hz, and between two and six DAS channels were used for cancelling the seismometer noise. The geophones measured ground velocity at a sampling rate of 400 Hz and each consisted of three orthogonal components: East, North, and Vertical. To ensure consistency with the seismometer data, the geophone and DAS recordings were both resampled to 200 Hz and subsequently detrended and normalized prior to analysis. Cross-spectral and auto-spectral densities were estimated using Daniell's method \cite{b15}.

\section{Results}
\subsection{DAS to Seismometer Conversion Results}
\begin{figure}[htbp]
	\centering
	\begin{minipage}{0.49\textwidth}
		\centering
		\includegraphics[width=\linewidth]{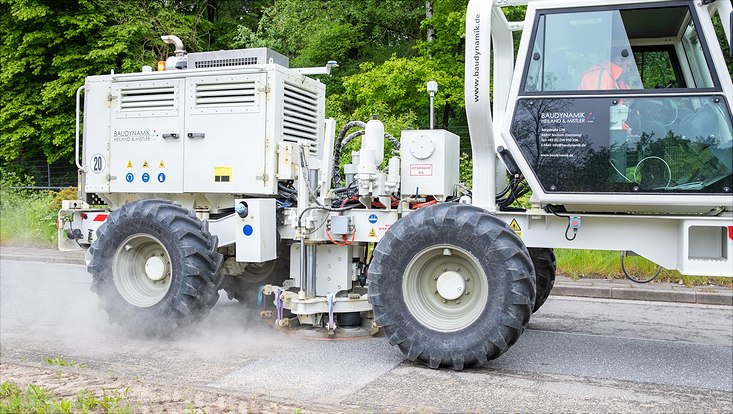}
	\end{minipage}
	\hfill
	\begin{minipage}{0.49\textwidth}
		\centering
		\includegraphics[width=\linewidth]{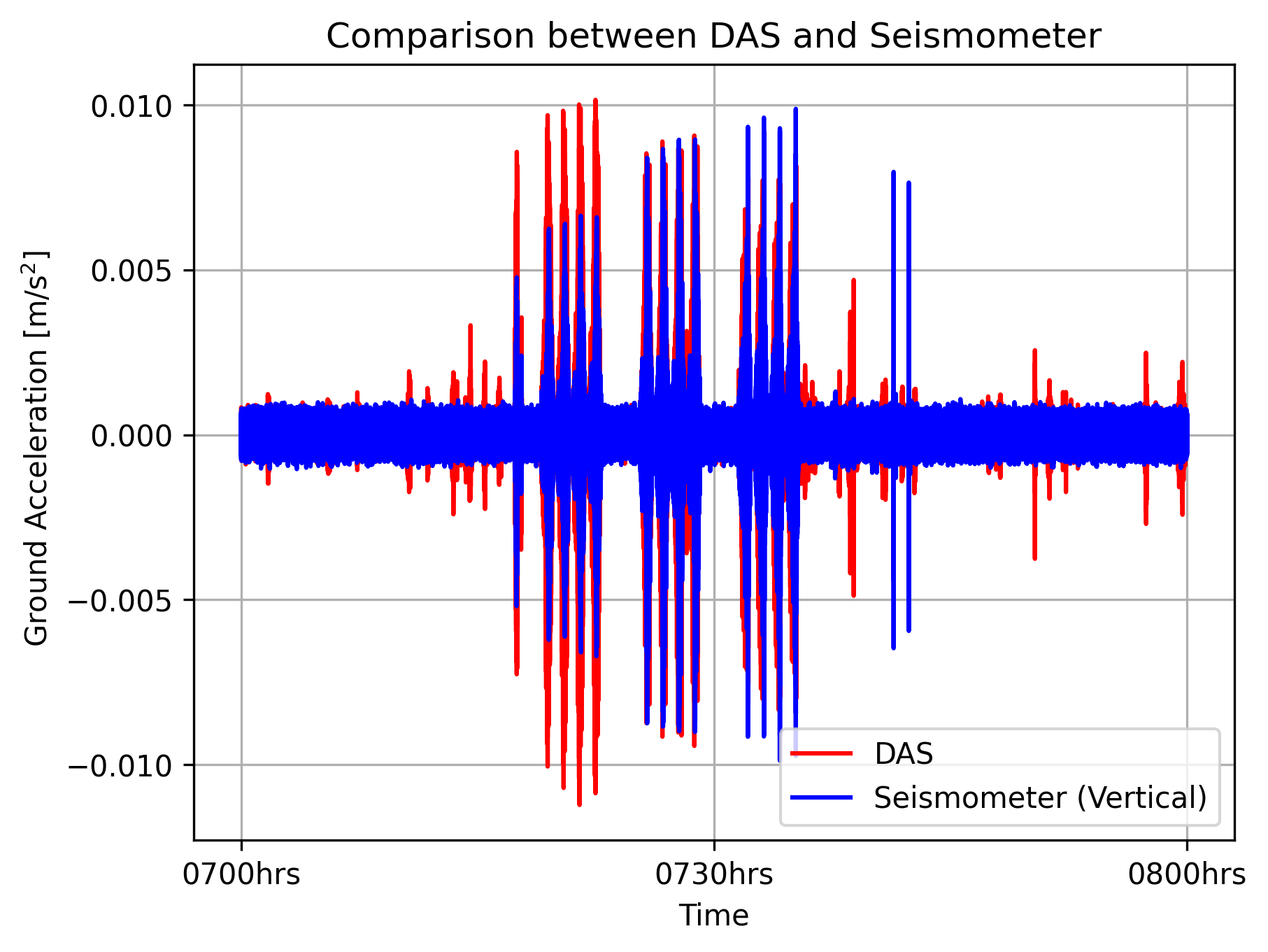}
	\end{minipage}
	\caption{\textit{Left: Vibrotruck used during the campus experiment to generate controlled ground motion. The vehicle excited the ground continuously for a duration of five minutes, providing a repeatable seismic source for comparison between DAS and seismometer recordings. Right: Time series comparison between the DAS output and the vertical component of a co-located seismometer. The alignment in time and waveform shape highlights the DAS system’s capability to accurately track seismic wave propagation.}}
	\label{fig6}
\end{figure}

From the experimental observations, it is evident that the output of DAS is comparable to that of a co-located broadband seismometer, with both instruments output converted to ground acceleration. To assess the level of agreement between the two systems, we analyzed their time-domain responses. Figure \ref{fig6} presents the waveform of the vibrotruck event recorded by both DAS and the seismometer. The alignment in onset time, signal amplitude, and overall waveform morphology indicates excellent consistency between the two measurements, confirming the capability of the DAS system to accurately capture seismic wave propagation.

\vspace{0.2cm}

\begin{figure}[H]
	\centering\includegraphics[width=18cm]{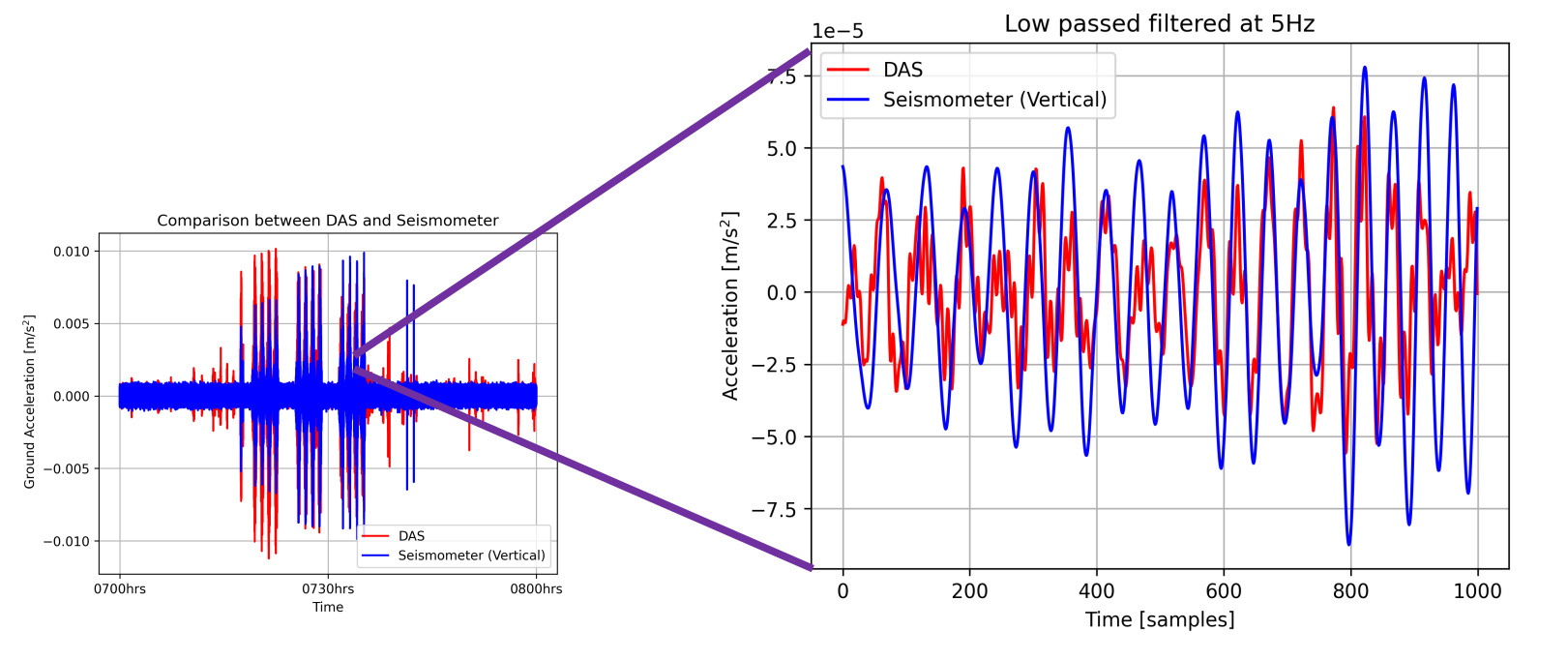}
	\caption{\textit{Comparison of DAS output with the vertical component of a co-located seismometer. The right panel shows a zoomed-in view of the vibrotruck event, where the oscillatory behavior is clearly visible in both recordings. The similarity in frequency content and waveform confirms the DAS system's capability to capture seismic signals with high fidelity.}}
	\label{fig7}
\end{figure}

\begin{figure}[htbp]
	\centering\includegraphics[width=18cm]{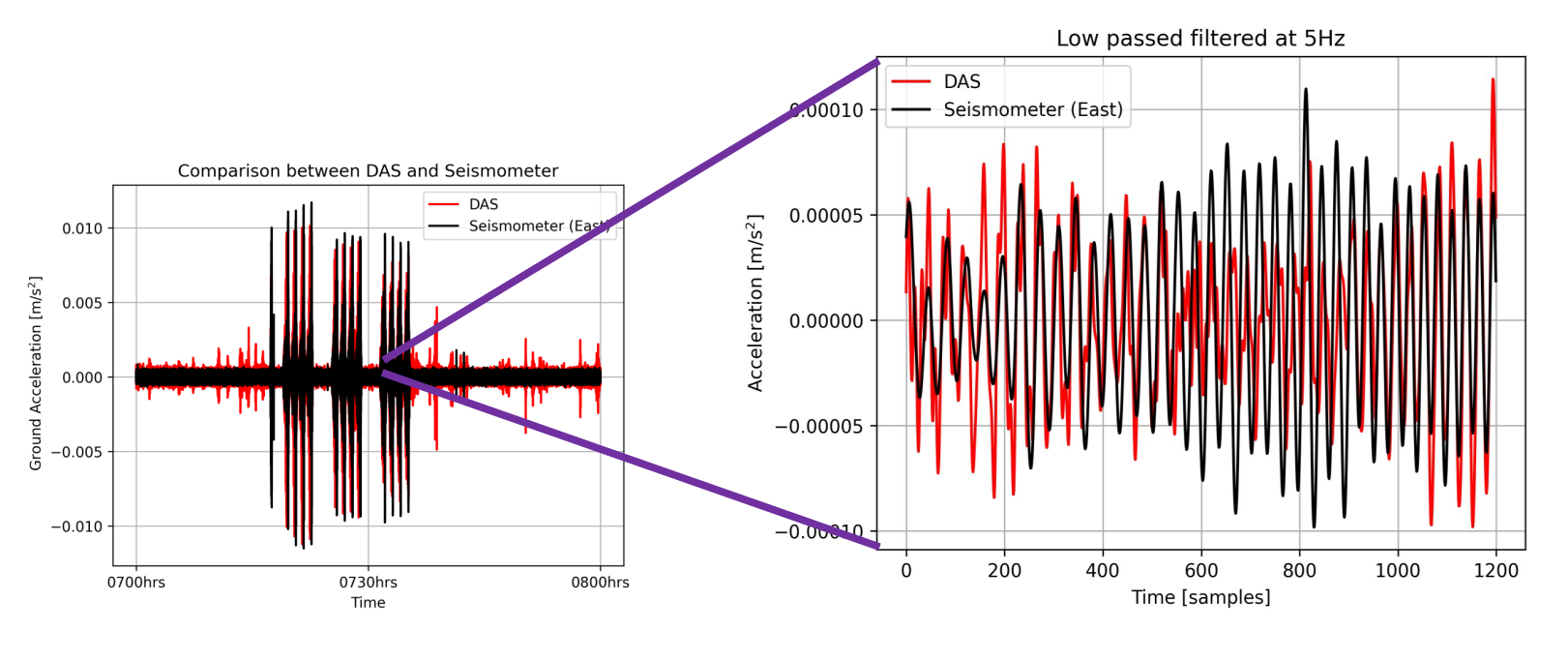}
	\caption{\textit{Comparison of DAS output with the east component of a co-located seismometer. The right panel shows a zoomed-in view of the vibrotruck event, where the oscillatory behavior is clearly visible in both recordings. The similarity in frequency content and waveform confirms the DAS system's capability to capture seismic signals with high fidelity.}}
	\label{fig8}
\end{figure}

\noindent Figure \ref{fig7} displays a comparison between the DAS output and the vertical component of the co-located seismometer. The right-hand panel provides a zoomed-in view, highlighting the oscillatory behavior of the vibrotruck event as captured by both sensors. The frequency content and waveform shape are clearly consistent across both instruments. A similar level of agreement is observed when comparing the DAS signal with the east component of the seismometer as shown in figure \ref{fig8}. These results demonstrate that DAS is capable of recording ground motion with a level of detail comparable to that of conventional seismometers, confirming its viability as a seismic sensing technology.

\vspace{0.2cm}

\begin{figure}[htbp]
	\centering
	\begin{minipage}{0.49\textwidth}
		\centering
		\includegraphics[width=\linewidth]{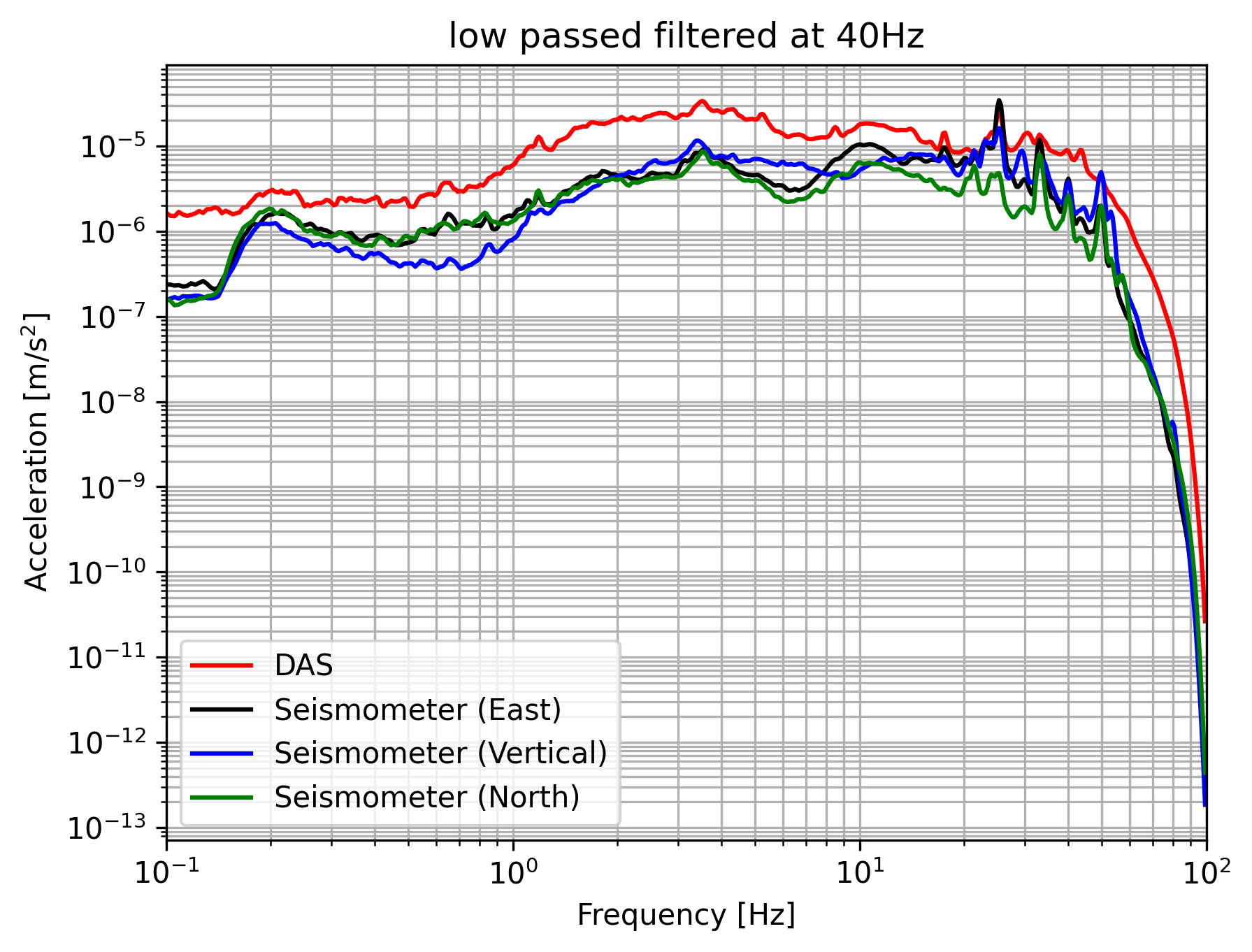}
	\end{minipage}
	\hfill
	\begin{minipage}{0.49\textwidth}
		\centering
		\includegraphics[width=\linewidth]{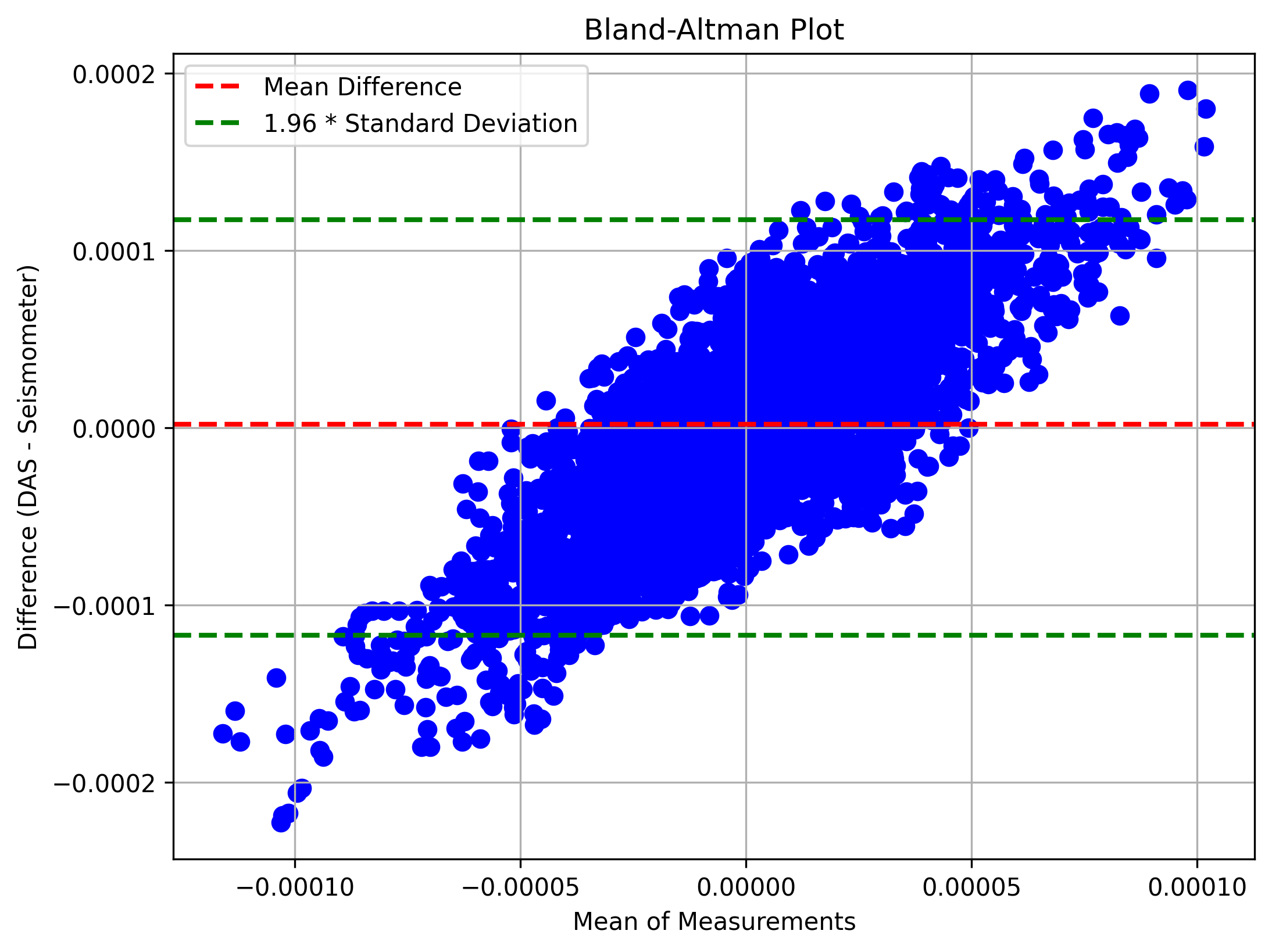}
	\end{minipage}
	\caption{\textit{Left: PSD comparison between the DAS output and the east, vertical, and north components of a co-located seismometer. The spectral shapes exhibit strong agreement, with the DAS showing higher amplitude due to spatial averaging, which enhances the SNR. Right: Bland–Altman plot assessing the agreement between DAS and the vertical component of the seismometer. The mean difference is centered around zero, and most data points lie within the ±1.96 standard deviation bounds, indicating good agreement between the two sensing systems.}}
	\label{fig9}
\end{figure}

\noindent To evaluate the spectral agreement between the DAS and a seismometer, we computed the power spectral density (PSD) of both sensors using LPSD \cite{b15} method. The DAS time series was averaged across 10 adjacent channels to improve the SNR. The analysis was performed on one hour of continuous time series data, low-pass filtered at 40 Hz, with a sampling rate of 200 Hz. Figure \ref{fig9} (left) shows the resulting PSD curves for both the averaged DAS output and the seismometer over the same time window. The spectral profiles are more similar across the 0.2–20 Hz frequency band, with both instruments capturing peaks corresponding to the ground motion induced by the vibrotruck excitation. Although the DAS exhibits slightly higher spectral amplitudes, this can be attributed to the spatial averaging process, which suppresses amplitude fluctuations due to uncorrelated noise and enhances coherent signal components.
The strong spectral similarity between the two systems confirms that DAS reliably captures the same frequency content as a traditional seismometer. These results support the suitability of DAS for applications such as Newtonian noise cancellation and general seismological monitoring, particularly in the frequency range of interest for ground-based gravitational wave detectors.

\vspace{0.2cm}
 
\noindent To further quantify the agreement between DAS and seismometer data, we employed a Bland–Altman applicable in metrology systems. This method plots the difference between the two measurements against their mean and provides insight into systematic biases or inconsistencies. As shown in figure \ref{fig9} (right), the differences between converted DAS and seismometer values are symmetrically distributed around zero, with no observable trend across the range of measurements. The mean difference lies within the zero limit, and the limits of agreement (mean ± 1.96 × standard deviation) encompass the majority of the data points. This result demonstrates that there is no bias between the two sensing types.

\vspace{0.2cm}

\noindent While some DAS channels exhibited high coherence with the seismometer, others—particularly those located in noisier or poorly coupled sections of the fiber—showed reduced SNR. Channels located near major infrastructure or experiencing loose coupling had visibly lower performance. This highlights the importance of fiber installation quality and channel selection in DAS deployment.

\subsection{Seismometer Noise Cancellation Using DAS and Geophone Arrays}
\noindent To evaluate the capability of DAS to cancel seismometer noise, we focus on the frequency range from 1 Hz to 20 Hz. We investigate the performance of noise cancellation for both the eastern and vertical components of the seismometer signal.
\vspace{0.2cm}

\noindent \textbf{Seismometer East Component Noise Cancellation with DAS}

\begin{figure}[htbp]
	\centering
	\begin{minipage}{0.49\textwidth}
		\centering
		\includegraphics[width=\linewidth]{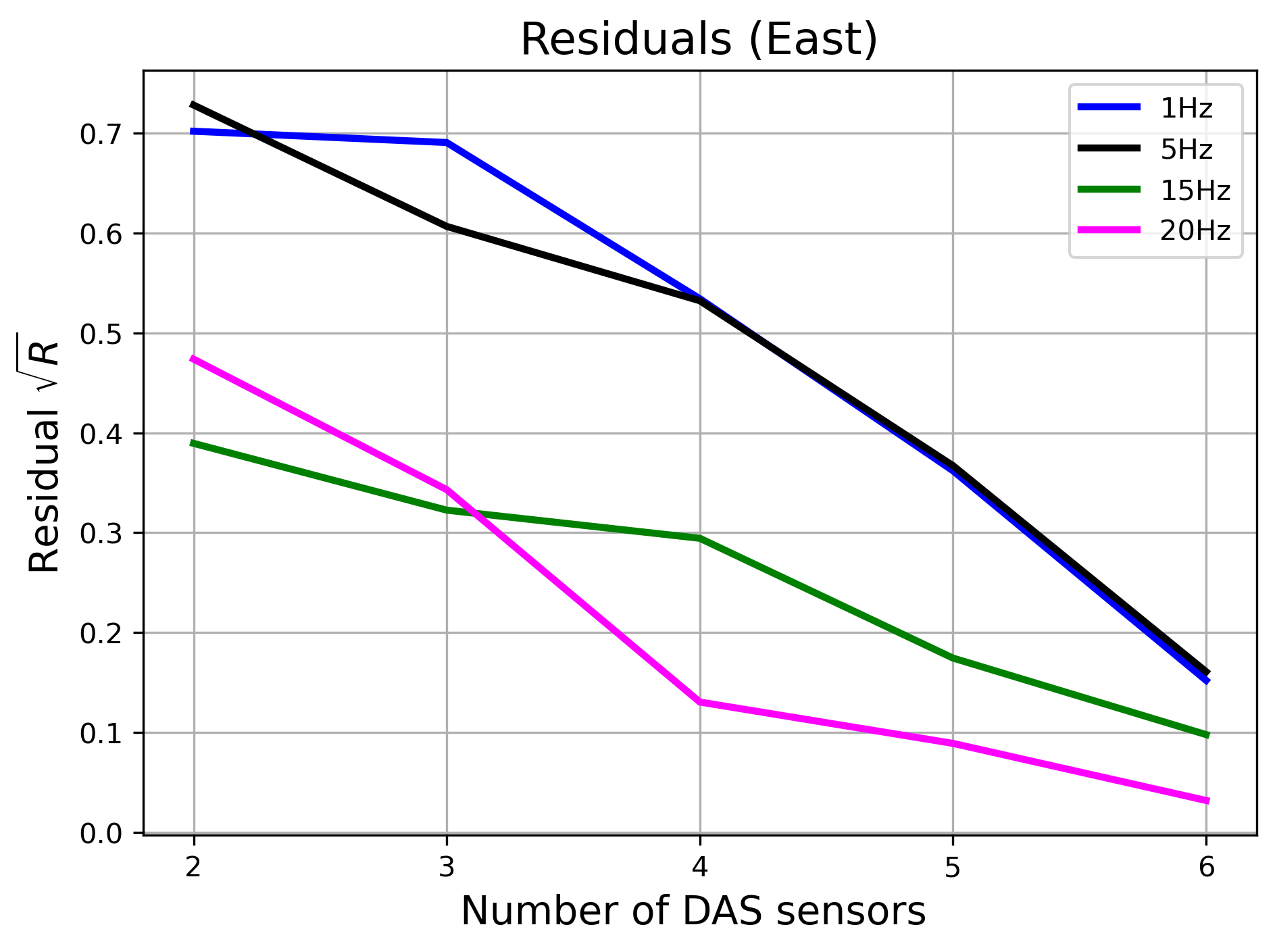}
	\end{minipage}
	\hfill
	\begin{minipage}{0.49\textwidth}
		\centering
		\includegraphics[width=\linewidth]{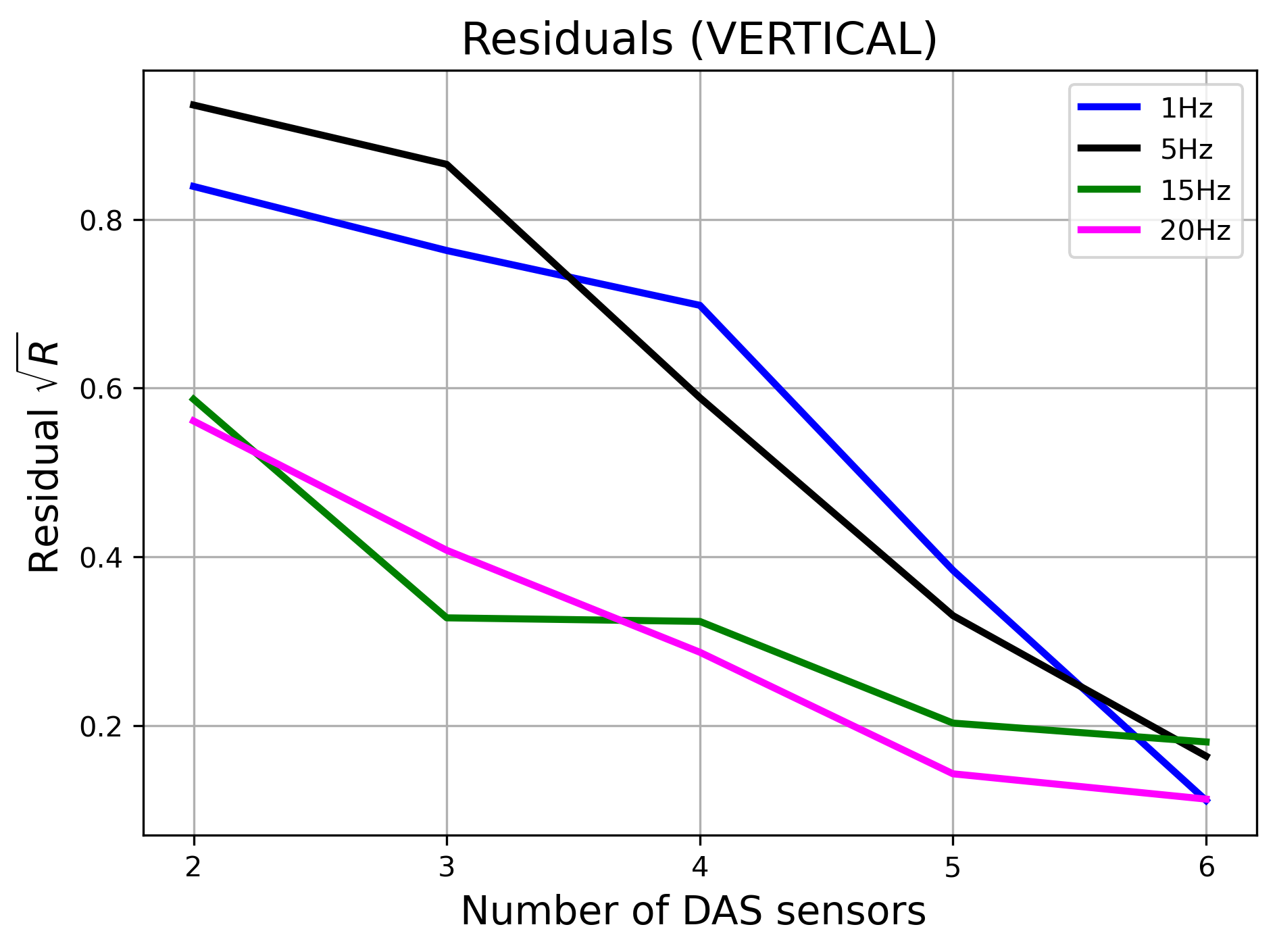}
	\end{minipage}
	\caption{\textit{Noise reduction factor as a function of the number of DAS channels used for cancelling seismometer noise.
			Left: Cancellation performance for the eastern component of the seismometer between 1 Hz and 20 Hz.
			Right: Cancellation performance for the vertical component of the seismometer over the same frequency range.
			The plots show that increasing the number of DAS channels  improves the noise reduction, with better performance observed for the eastern component. This is attributed to the alignment of the DAS sensing axis with the horizontal motion measured by the seismometer.}}
	\label{fig69}
\end{figure}

\noindent We observe that using two DAS sensors results in a noise reduction factor of 0.70 at 1 Hz. As the number of DAS channels increases, the reduction factor improves significantly: using six DAS sensors, the reduction factor reaches 0.15. At 20 Hz, the DAS system achieves a reduction factor of 0.47 with two channels, improving to 0.04 when six DAS sensors are utilized as demonstrated in figure \ref{fig69}. These results shows that DAS is effective in cancelling the eastern component of the seismometer noise, especially with increasing the number of channels.

\vspace{0.2cm}

\noindent \textbf{Seismometer Vertical Component Noise Cancellation with DAS}

\noindent We observe that at 1 Hz, the noise can be reduced by a factor of 0.82 when using two DAS channels. Increasing the number of DAS sensors to six leads to a further reduction factor of 0.14. At 20 Hz, the cancellation factor is 0.58 with two DAS channel, decreasing to 0.14 when six channels are used. These findings confirm the feasibility of using DAS to cancel vertical seismometer noise as well.

\vspace{0.2cm}

\noindent Comparing the two components, we note that the cancellation factors are consistently lower (i.e., better cancellation) for the eastern component compared to the vertical component. This difference is attributed to the alignment of the DAS measurement axis: the DAS sensors primarily recorded strain rate along the horizontal axis, which closely matches the measurement direction of the eastern component of the seismometer. In contrast, the vertical component cancellation was less effective due to the mismatch between the strain measurement axis of the DAS system and the vertical motion measured by the seismometer.

\subsubsection{Comparison of Seismometer Noise Cancellation Using DAS and Geophones}
\noindent Geophones have been previously employed in Newtonian noise cancellation efforts at KAGRA and Virgo \cite{b14}. To prove the capability of DAS for Newtonian noise cancellation, we directly compared its performance with that of a traditional geophone array.  In this study, we used three triaxial geophones (totaling nine channels), from which we selected six channels—three East and three Vertical components—to match six DAS channels for a direct comparison. These selected channels were used to cancel the East and Vertical components of a colocated seismometer.

\vspace{0.2cm}

\begin{figure}[htbp]
	\centering
	\begin{minipage}{0.49\textwidth}
		\centering
		\includegraphics[width=\linewidth]{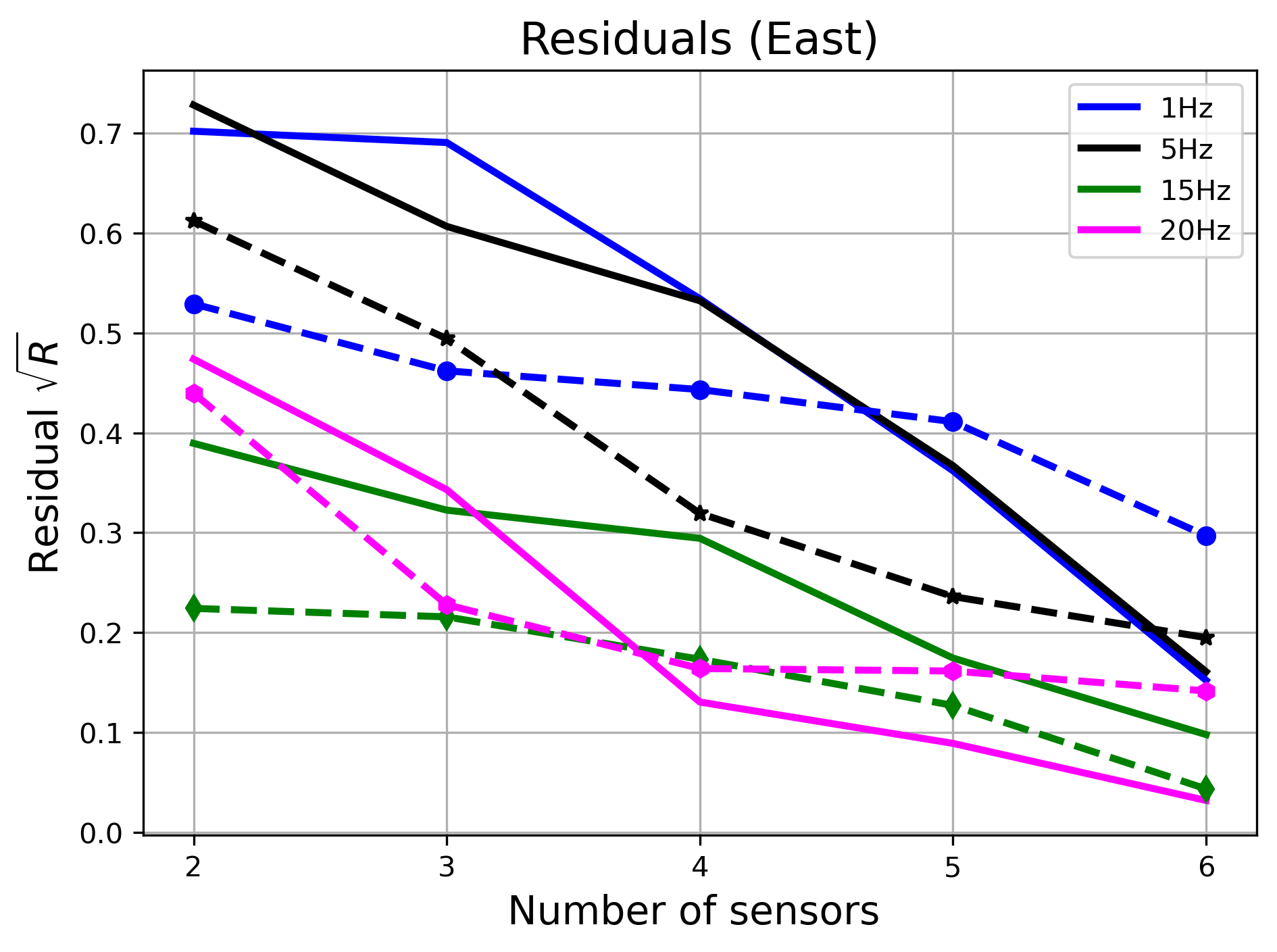}
	\end{minipage}
	\hfill
	\begin{minipage}{0.49\textwidth}
		\centering
		\includegraphics[width=\linewidth]{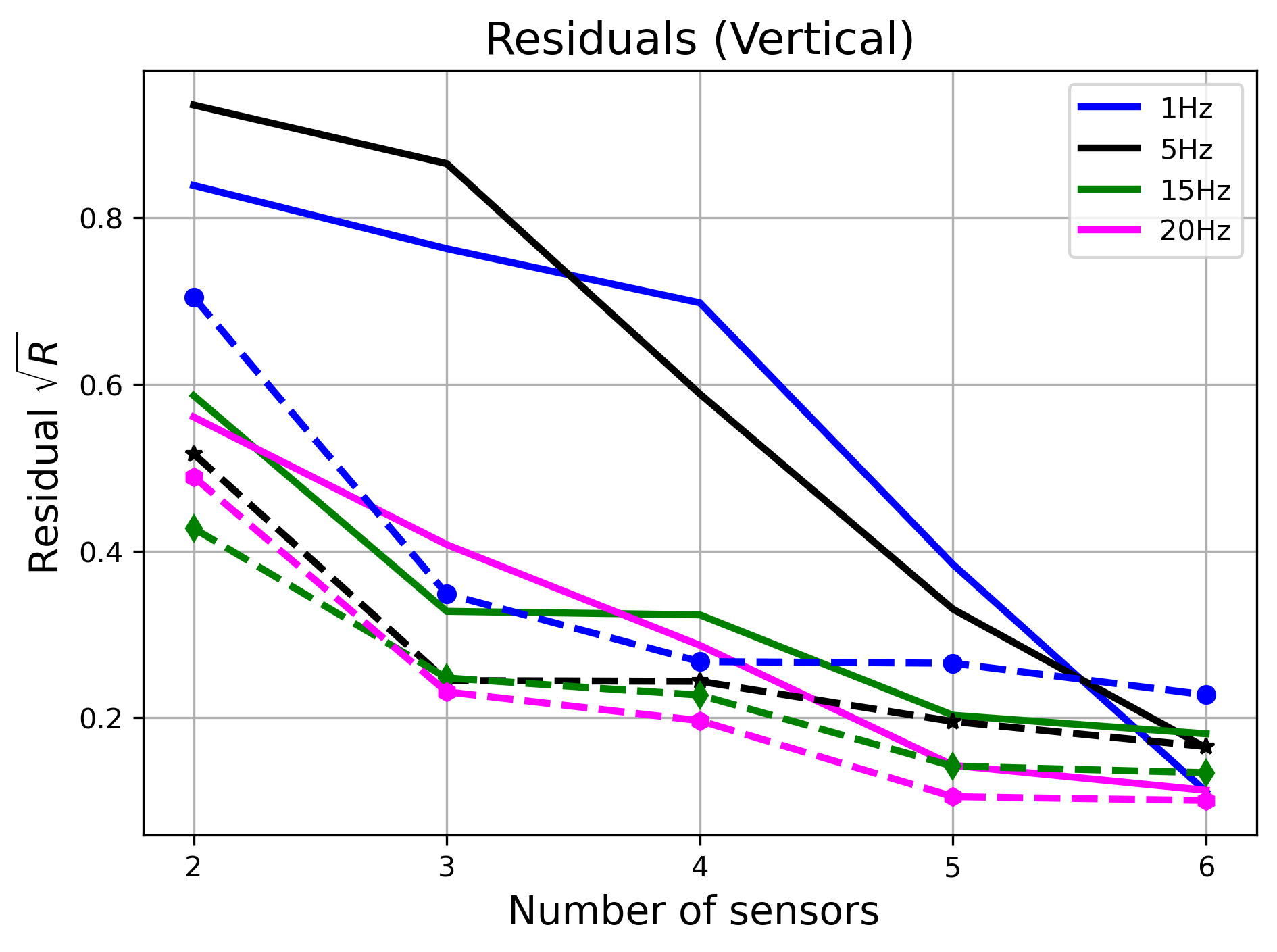}
	\end{minipage}
	\caption{\textit{Noise reduction factor comparison between geophones and DAS as a function of the number of sensors used, geophones are the dashed with the symbol plots.
			Left: Cancellation performance for the eastern component of the seismometer between 1 Hz and 20 Hz.
			Right: Cancellation performance for the vertical component of the seismometer over the same frequency range.
			The plots show that increasing the number of sensors channels significantly improves the noise reduction, with better performance observed for the eastern component.}}
	\label{fig99}
\end{figure}

\noindent \textbf{East Component Cancellation}
\\
\noindent At 5 Hz, cancellation using two geophone channels yielded a residual of 0.62, while using two DAS channels resulted in a residual of 0.73. Increasing the number of channels to six improved the performance significantly for both systems: the geophone array achieved a residual of 0.20, while DAS outperformed with a lower residual of 0.17. At 20 Hz, the geophone-based cancellation gave a residual of 0.43 with two channels and 0.15 with six channels. In comparison, DAS achieved a residual of 0.47 with two channels and a notably better residual of 0.04 with six channels. A summary of the residuals across frequencies is presented in figure \ref{fig99} and summarized in table \ref{tab:nn_cancellation_residuals}.

\vspace{0.2cm}

\noindent \textbf{Vertical Component Cancellation:}
\\
\noindent When targeting the vertical component at 5 Hz using two channels, the geophone array produced a residual of 0.52, while DAS recorded a higher residual of 0.88. However, with six channels, both systems converged to the same residual of 0.16. At 20 Hz, two geophone channels achieved a residual of 0.50, while DAS yielded 0.58. With six channels, the geophone array reduced the residual to 0.08, whereas DAS achieved 0.14.

\vspace{0.2cm}

\noindent These findings demonstrate that DAS-based seismometer  cancellation is comparable in performance to that achieved with geophone arrays. In particular, DAS performed better when the number of channels was increased. Future work should explore hybrid configurations where DAS systems are co-optimized with seismometers and geophones to enhance Newtonian noise cancellation performance.
\begin{table}[ht]
	\centering
	\caption{Residuals of Seismometer Noise Cancellation Using DAS and Geophones}
	\begin{tabular}{|c|c|c|c|c|}
		\hline
		\textbf{Component} & \textbf{Frequency (Hz)} & \textbf{Channels Used} & \textbf{Geophone Residual} & \textbf{DAS Residual} \\
		\hline
		
		\multirow{10}{*}{East} 
		& \multirow{5}{*}{5}  & 2 & 0.62 & 0.73 \\
		&                     & 3 & 0.50 & 0.61 \\
		&                     & 4 & 0.32 & 0.55 \\
		&                     & 5 & 0.24 & 0.37 \\
		&                     & 6 & 0.20 & 0.17 \\
		\cline{2-5}
		& \multirow{5}{*}{20} & 2 & 0.43 & 0.47 \\
		&                     & 3 & 0.22 & 0.33 \\
		&                     & 4 & 0.17 & 0.14 \\
		&                     & 5 & 0.16 & 0.09 \\
		&                     & 6 & 0.15 & 0.04 \\
		\hline
		
		\multirow{15}{*}{Vertical} 
		& \multirow{5}{*}{5}   & 2 & 0.52 & 0.88 \\
		&                      & 3 & 0.22 & 0.86 \\
		&                      & 4 & 0.23 & 0.59 \\
		&                      & 5 & 0.20 & 0.34 \\
		&                      & 6 & 0.16 & 0.16 \\
		\cline{2-5}
		& \multirow{5}{*}{15}  & 2 & 0.49 & 0.59 \\
		&                      & 3 & 0.22 & 0.36 \\
		&                      & 4 & 0.20 & 0.35 \\
		&                      & 5 & 0.10 & 0.20 \\
		&                      & 6 & 0.10 & 0.18 \\
		\cline{2-5}
		& \multirow{5}{*}{20}  & 2 & 0.42 & 0.58 \\
		&                      & 3 & 0.22 & 0.40 \\
		&                      & 4 & 0.20 & 0.30 \\
		&                      & 5 & 0.10 & 0.16 \\
		&                      & 6 & 0.11 & 0.11 \\
		\hline
	\end{tabular}
	\label{tab:nn_cancellation_residuals}
\end{table}

\vspace{0.2cm}

\begin{figure}[htbp]
	\centering
	\begin{minipage}{0.5\textwidth}
		\centering
		\includegraphics[width=\linewidth]{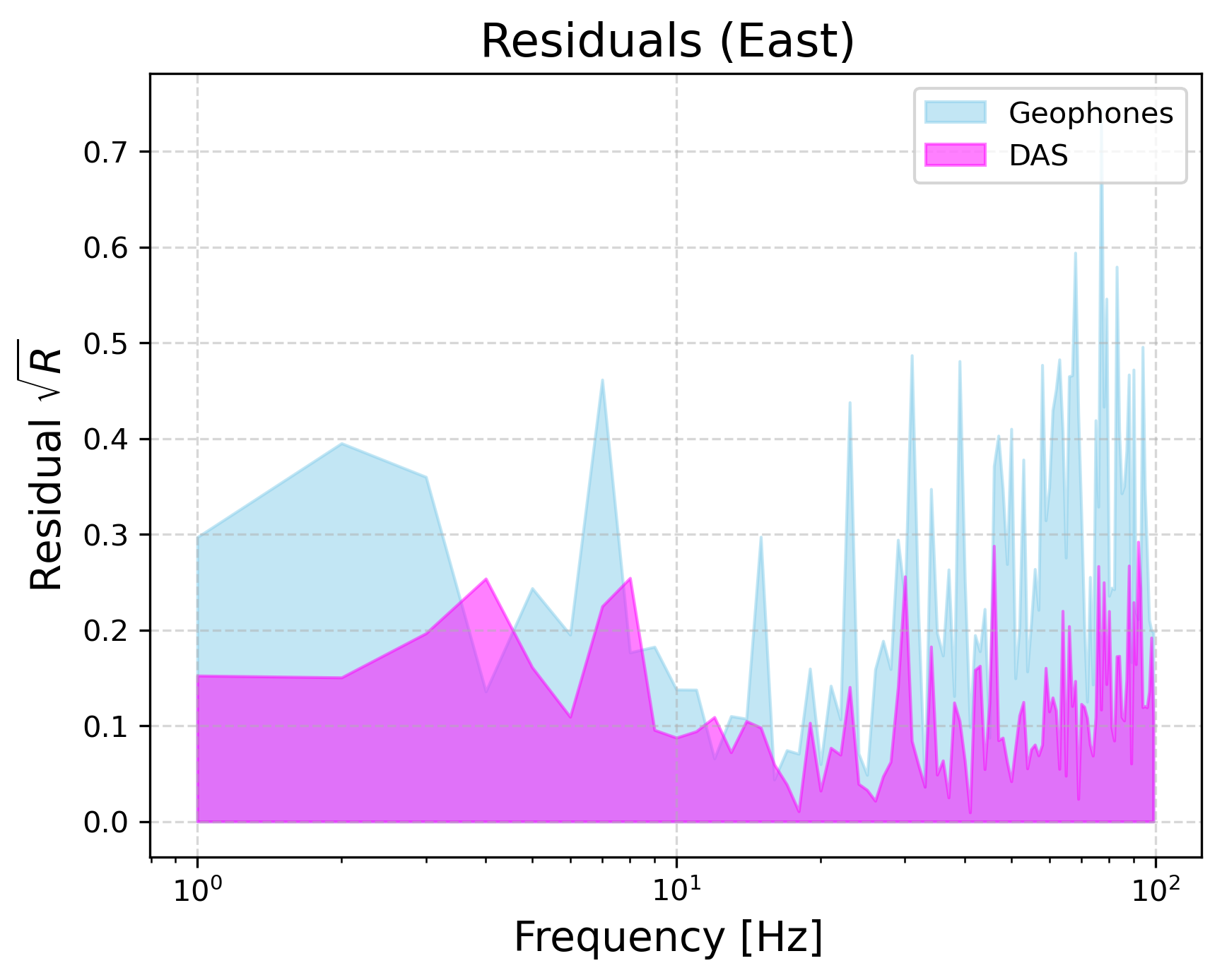}
	\end{minipage}
	\hfill
	\begin{minipage}{0.49\textwidth}
		\centering
		\includegraphics[width=\linewidth]{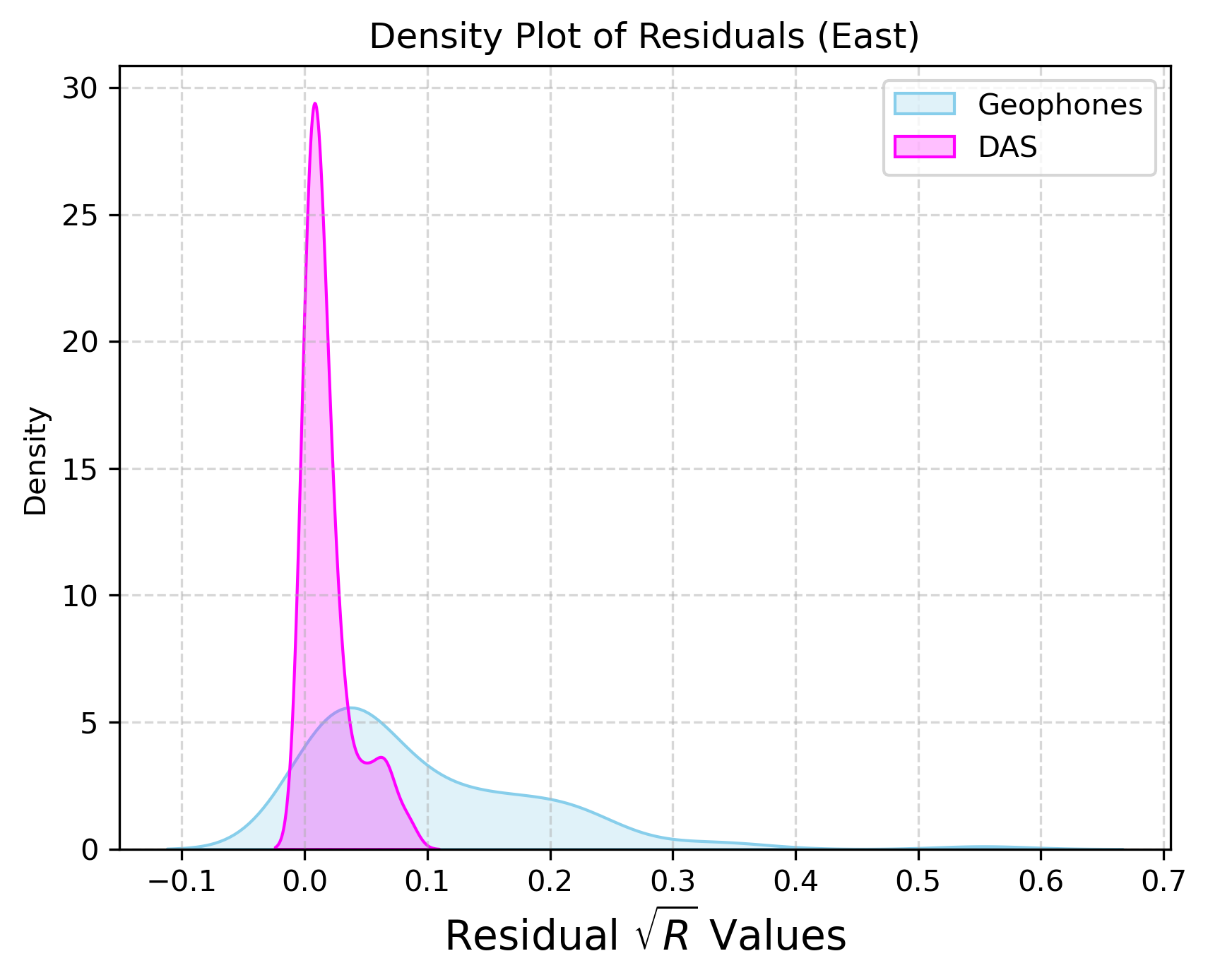}
	\end{minipage}
	\vspace{0.3cm}
	\begin{minipage}{0.5\textwidth}
		\centering
		\includegraphics[width=\linewidth]{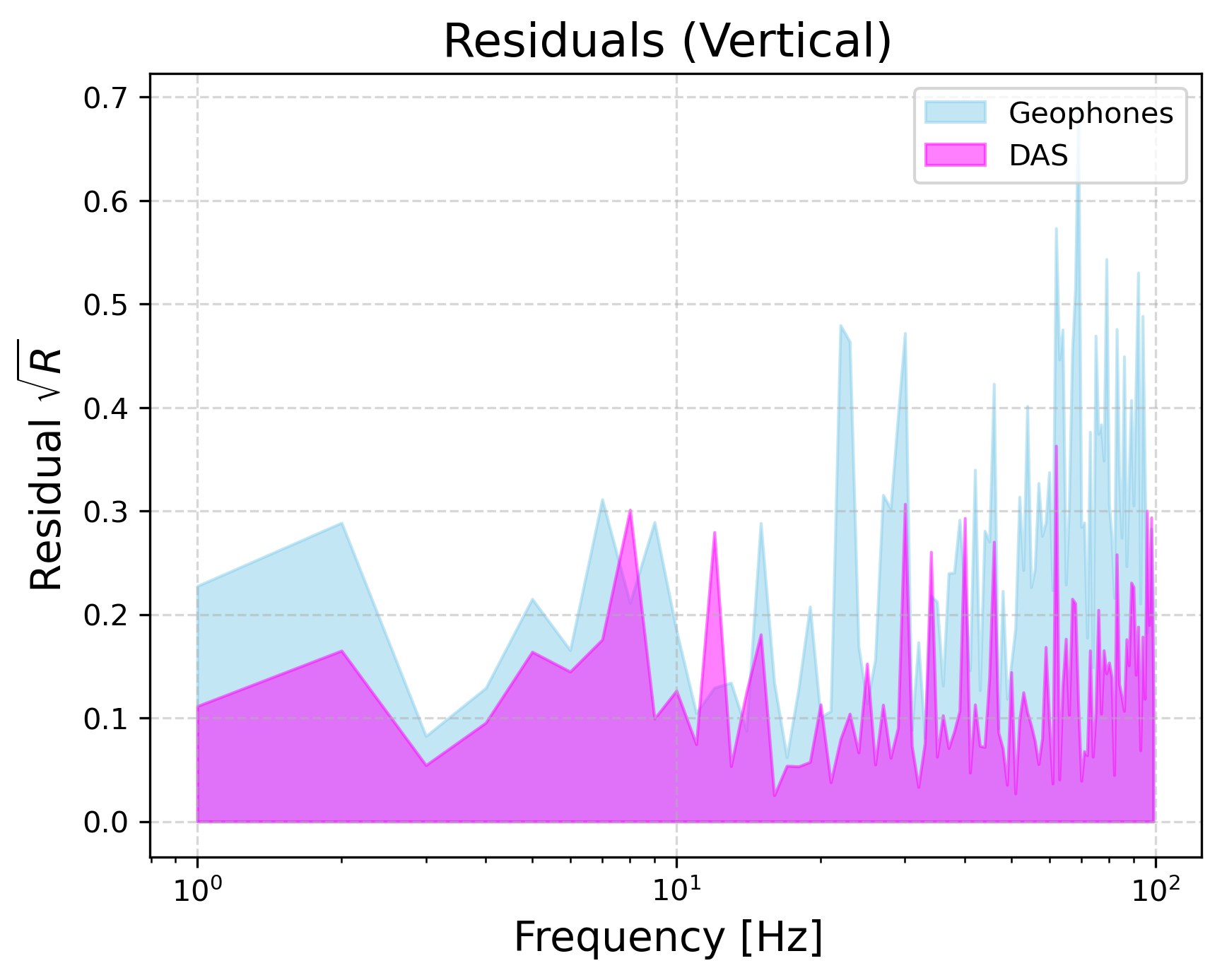}
	\end{minipage}
	\hfill
	\begin{minipage}{0.49\textwidth}
		\centering
		\includegraphics[width=\linewidth]{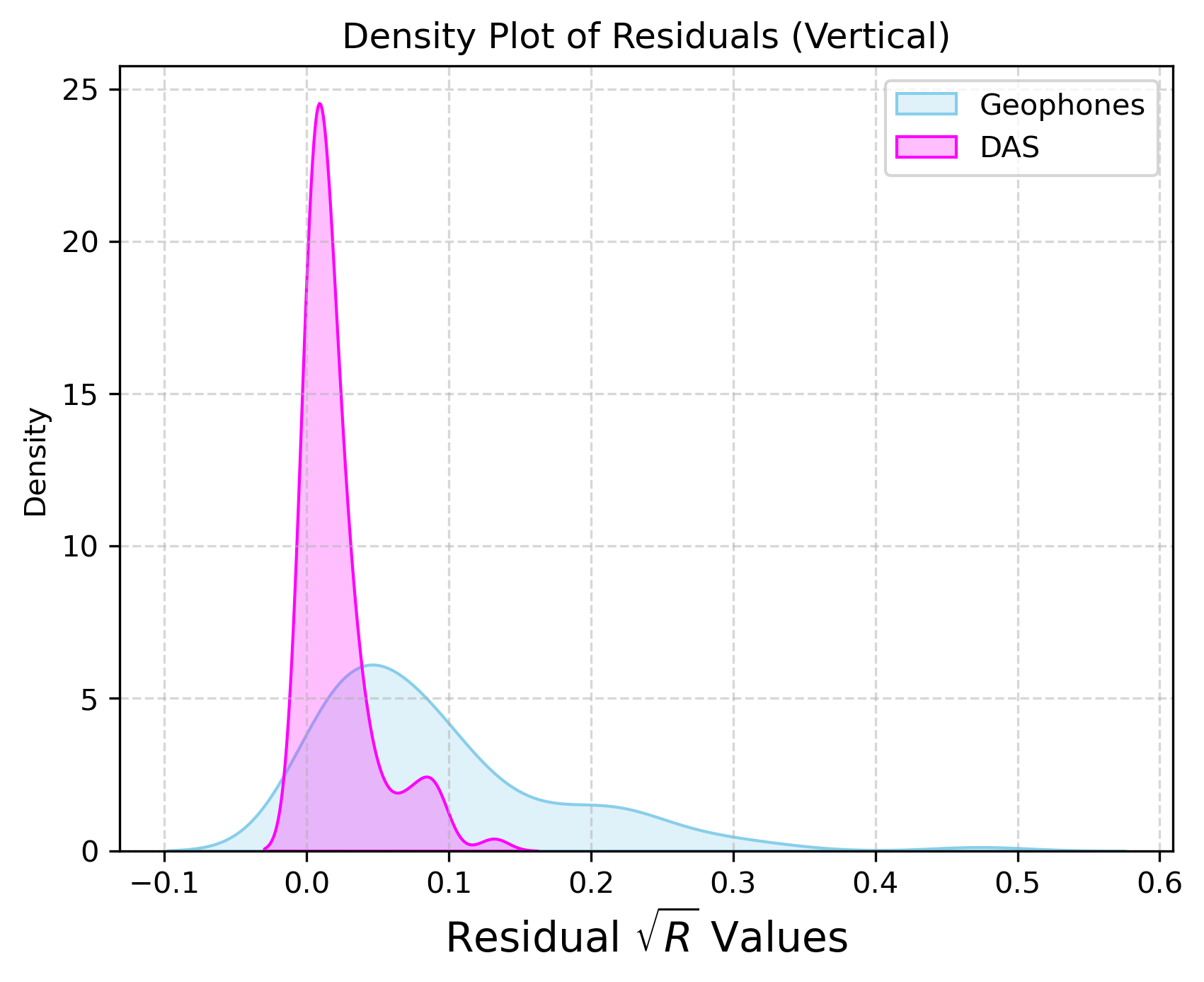}
	\end{minipage}
	\caption{\textit{Residuals plotted as a function of frequency for the noise cancellation process when 6 channels are used. The PDF of the DAS residuals shows a sharp peak near zero, indicating that most residuals are small and concentrated around zero. The peak occurs between 0.00 and 0.02 for both the vertical and eastern components of the seismometer. Geophones distribution is right-skewed, with a long tail extending toward larger residuals. The lack of multimodality  between the two sensor types suggests a consistent and effective noise cancellation behavior.}}
	\label{fig70}
\end{figure}

\noindent Figure \ref{fig70} presents the comparisons between DAS and geophone residuals plotted as a function of frequency in the range from 1 Hz to 99 Hz. We observe varying values of the residual factors across frequencies. To better understand the distribution of these residuals, we constructed a probability density function (PDF) plot. DAS's PDF plot reveals a sharp peak close to zero, indicating that the majority of residuals ($\sqrt{R}$ values) are small and tightly clustered near zero. Specifically, we observe a high density, peaking around 30, with the maximum located between 0.00 and 0.02 for both the vertical and eastern components when using DAS as witness sensors.

\vspace{0.2cm}

\noindent We observe a broad PDF when using geophones as witness sensors. The distribution is right-skewed, with a longer tail extending toward positive residual values. This suggests that while most residuals are very small, a few larger values occur but are relatively rare. With both DAS and geophones, there are no multimodalities observed which indicate a single dominant cancellation behavior. These results further demonstrate that DAS is  effective at cancelling the seismometer noise, as reflected by the concentration of residuals near zero and its results are comparable or to geophones.

\vspace{0.2cm}

\noindent The experimental results presented in this study demonstrate that DAS effectively cancels seismometer noise, particularly in the vertical component, challenging the claims made by the authors of \cite{b2a}. Specifically, the authors of \cite{b2a} argued that strain-meters (for example, DAS) must be deployed far from the gravitational wave detector test mass to achieve a strong correlation between the test mass displacement (or acceleration) and the strain-meter measurements. However, the experimental data from this study contradict that theoretical expectation.

\vspace{0.2cm}

\noindent Our results show that when DAS and a seismometer measure displacement (acceleration) along the same axis, a high correlation is achieved when the sensors are colocated. Furthermore, our  study shows that placing DAS sensors far from the seismometer degrades the local signal's SNR. This degradation arises because DAS signal processing involves spatial averaging, which incorporates signals from a broader region, including those unrelated to the local measurement. Nevertheless, DAS can effectively record longer-wavelength signals by averaging the output over spatial extents larger than the wavelength of the target signal.

\vspace{0.2cm}

\noindent These findings highlight the need for further investigation, particularly regarding the true displacement of gravitational wave detector test masses and how closely DAS measurements can track these displacements.

\subsection{Geophone Data Estimation}
The Wiener filter successfully estimated the geophone data from the DAS measurements, demonstrating the potential of DAS systems for seismic signal reconstruction. The correlation coefficient between the estimated and true geophone signals averaged 0.97, indicating a strong linear relationship between the two. The amplitude spectra of the DAS-derived geophone estimates closely matched the spectra of the true geophone data. the Wiener filter was capable of preserving the temporal characteristics of the seismic events, further confirming the filter's effectiveness in estimating the true geophone signal.

\vspace{0.2cm}

\noindent These results show that DAS is capable of estimating geophone data under the conditions tested in this study. This finding  suggests that DAS can be used as a cost-effective alternative or supplement to traditional seismic networks, particularly in environments where the deployment of geophones or seismometers is impractical or expensive. By leveraging the spatially dense and continuous nature of DAS arrays, it is possible to extend seismic monitoring capabilities across larger areas with fewer physical sensors, without sacrificing the quality of the data.

\begin{figure}[htbp]
	\centering
	\begin{minipage}{0.49\textwidth}
		\centering
		\includegraphics[width=\linewidth]{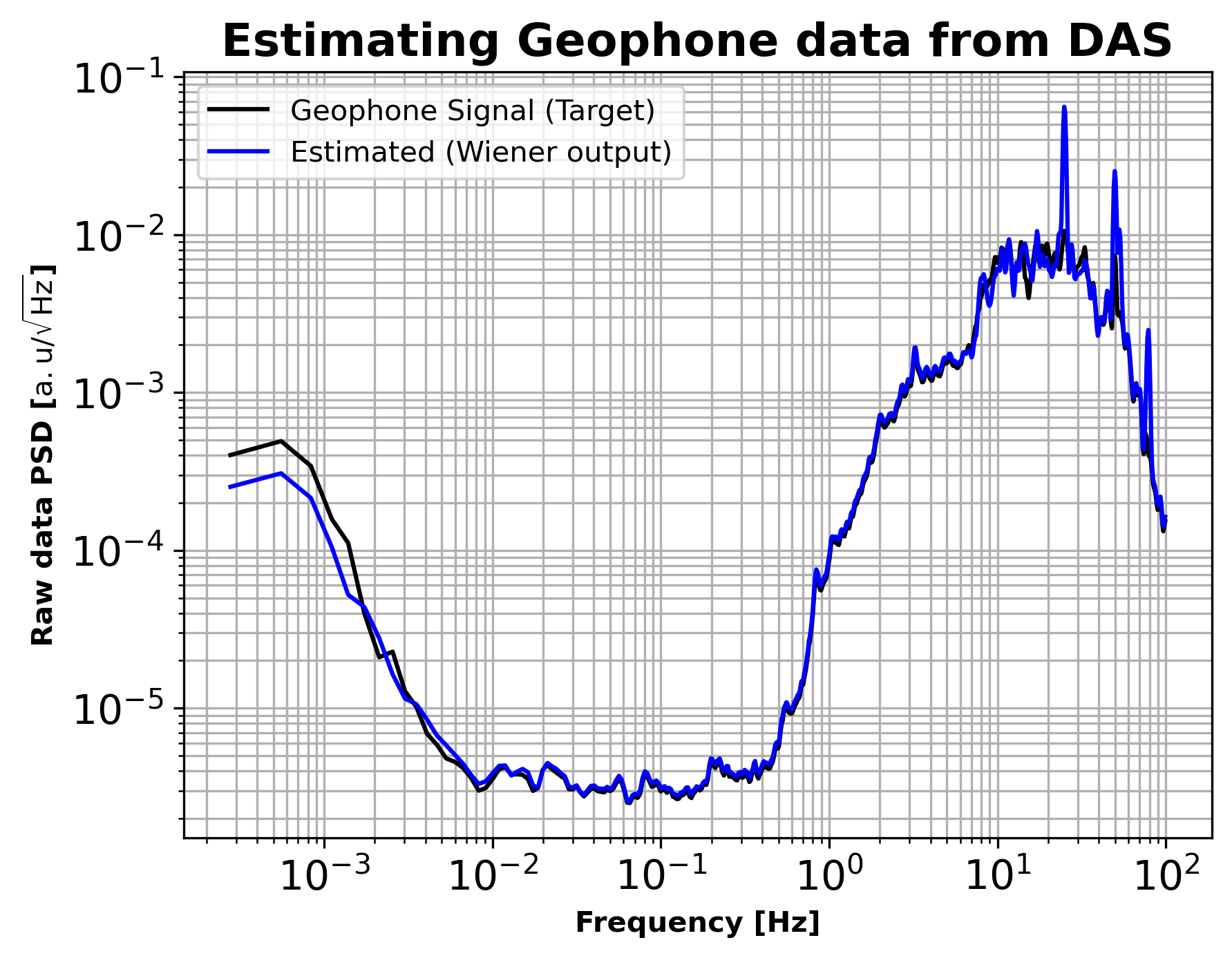}
	\end{minipage}
	\hfill
	\begin{minipage}{0.49\textwidth}
		\centering
		\includegraphics[width=\linewidth]{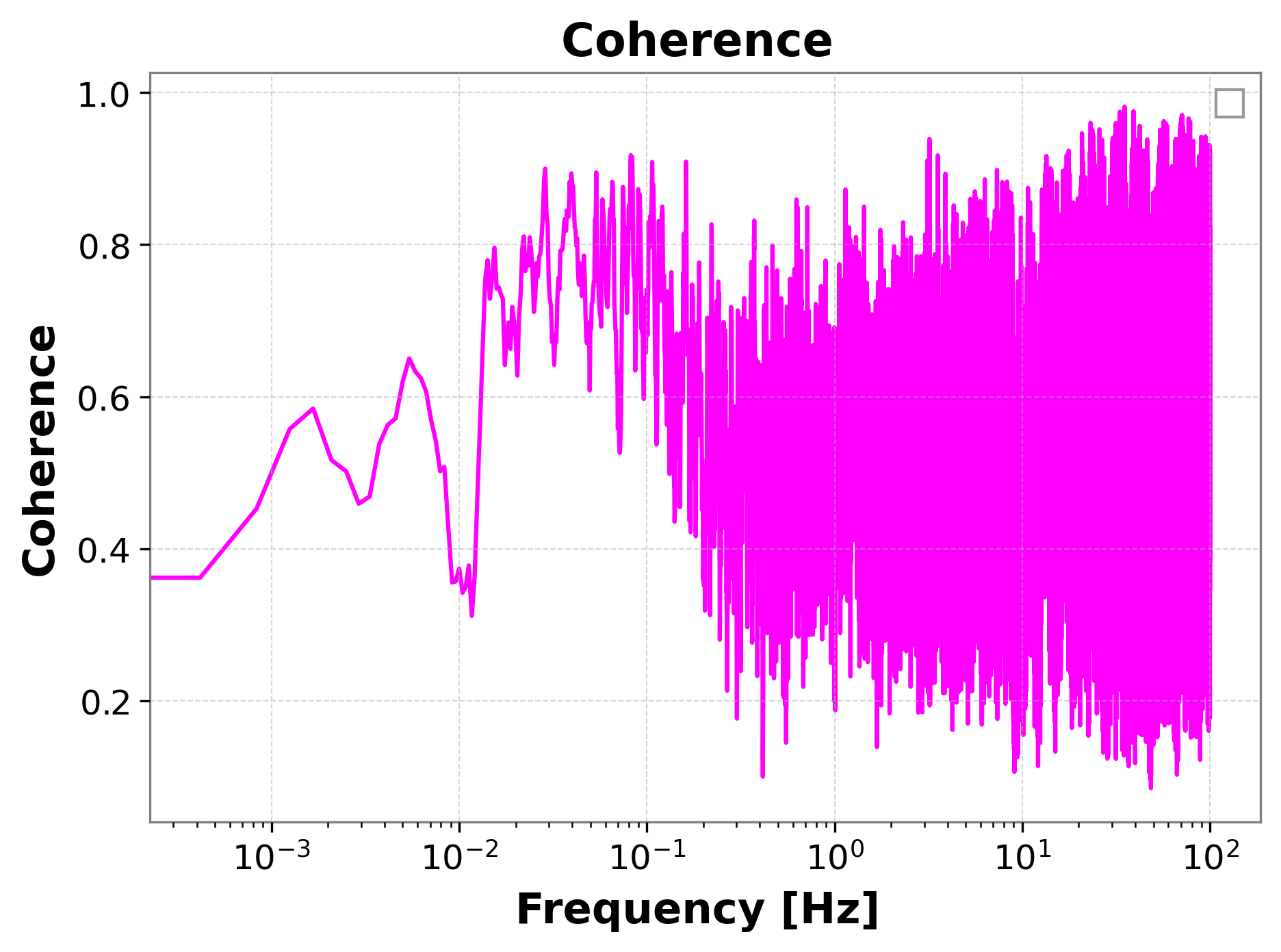}
	\end{minipage}
	\caption{\textit{Comparison of DAS-predicted geophone signals with actual geophone data. Left: PSD plot showing strong agreement across all frequencies between the Wiener filter output and the real geophone signal. Right: Coherence remains above 0.7 for frequencies above 3 Hz, indicating high correlation between the predicted and measured geophone data.}}
	\label{fig199}
\end{figure}

\section{DAS Coherence Length}
The LIGO Hanford Observatory experiences non-coherent ground motion at frequencies relevant for Newtonian Noise mitigation, especially near the optics. This presents a challenge for traditional seismic arrays with sparse spacing (e.g., 25 meters), which may fail to resolve local ground motion features. To address this, we investigated the coherence length of DAS during a controlled vibrotruck event in the 3–10 Hz frequency range

\subsubsection{3–5 Hz Frequency Band}
\begin{figure}[htbp]
	\centering
	\begin{minipage}{0.49\textwidth}
		\centering
		\includegraphics[width=\linewidth]{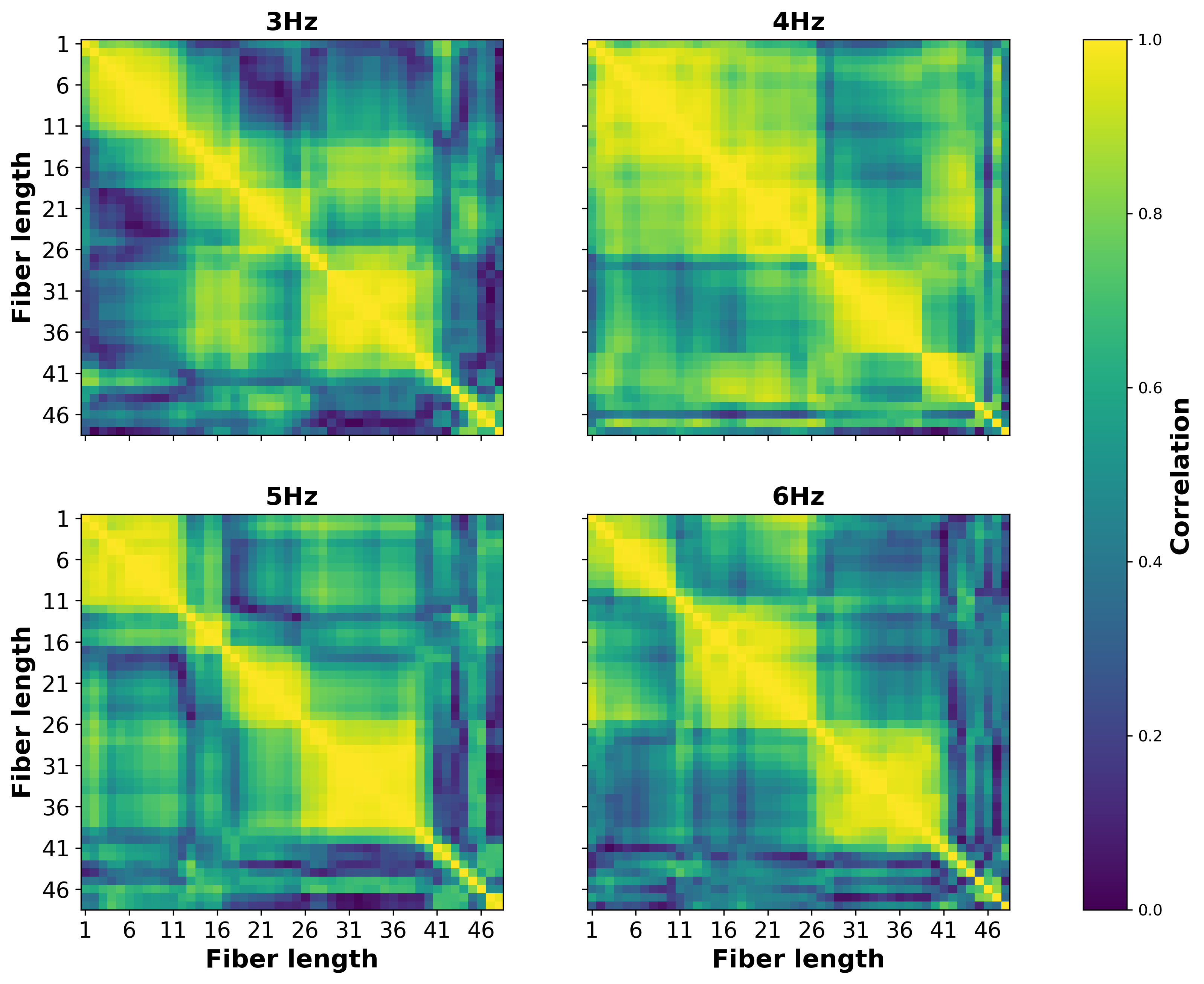}
	\end{minipage}
	\hfill
	\begin{minipage}{0.49\textwidth}
		\centering
		\includegraphics[width=\linewidth]{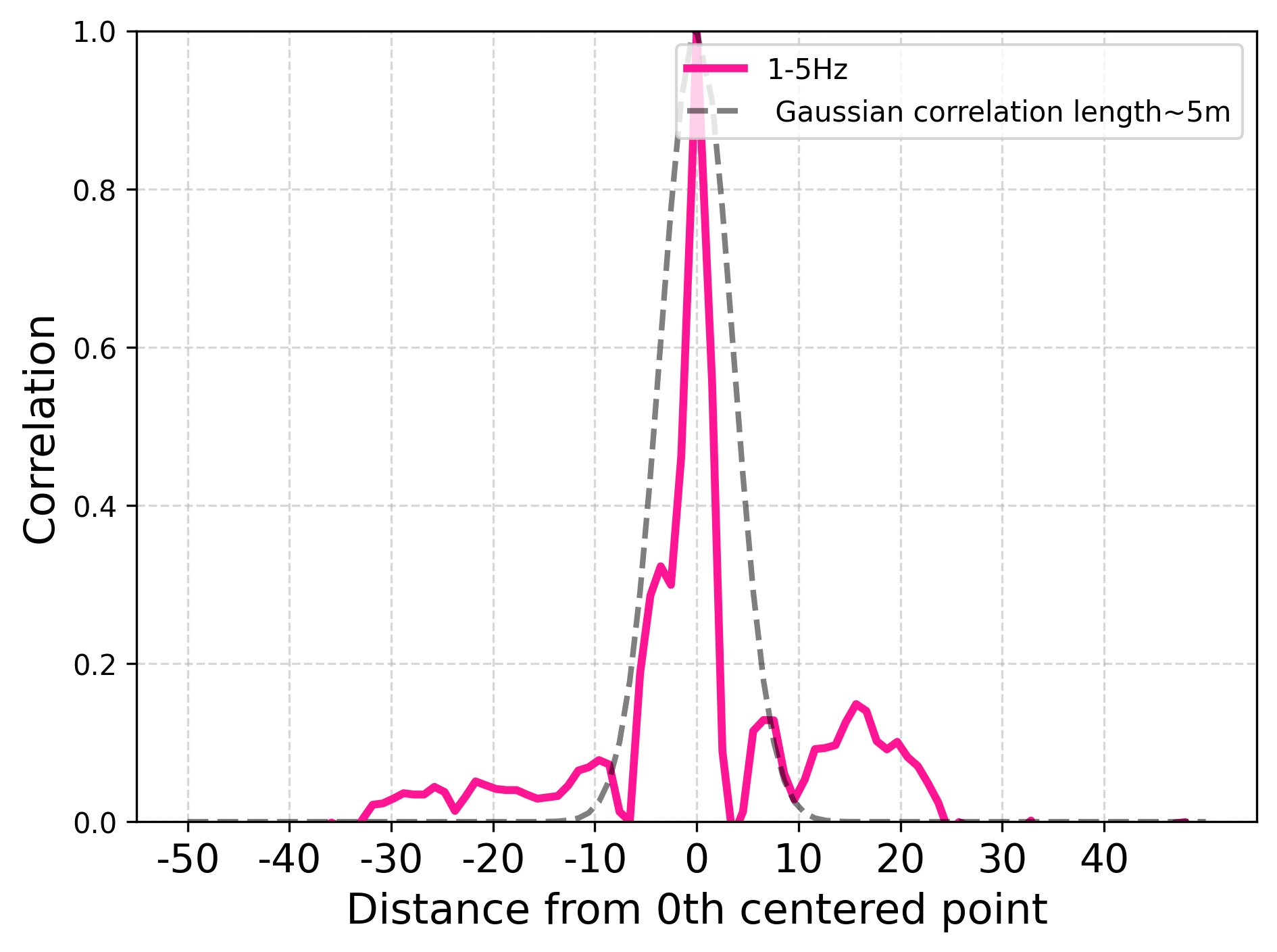}
	\end{minipage}
	\caption{\textit{Coherence length analysis of DAS in the 3–6 Hz band. Left: Coherence plots at 3, 4, 5, and 6 Hz showing a consistent coherence length of approximately 11 meters across the DAS array. Right: Comparison between the DAS-derived correlation profile and the corresponding Gaussian model, yielding a Gaussian correlation length of 11.78 meters, in good agreement with the DAS coherence length.}}
	\label{fig299}
\end{figure}

In the 3–5 Hz range, we observe that the DAS system exhibits a spatial coherence length of approximately 11 meters, as illustrated in figure \ref{fig299}. For validation, we compared this value to the Gaussian correlation length derived from fitting the spatial correlation function. The Gaussian correlation length in this band is found to be 11.76 meters, which closely matches the DAS-derived coherence length. The Gaussian coherence length is calculated using the relation:
\begin{equation}
	L_{coh}=2.355 \cdot L_{c},
\end{equation} where $L_{c}$ is the standard deviation of the Gaussian fit.

\subsubsection{6–10 Hz Frequency Band}
\begin{figure}[htbp]
	\centering
	\begin{minipage}{0.49\textwidth}
		\centering
		\includegraphics[width=\linewidth]{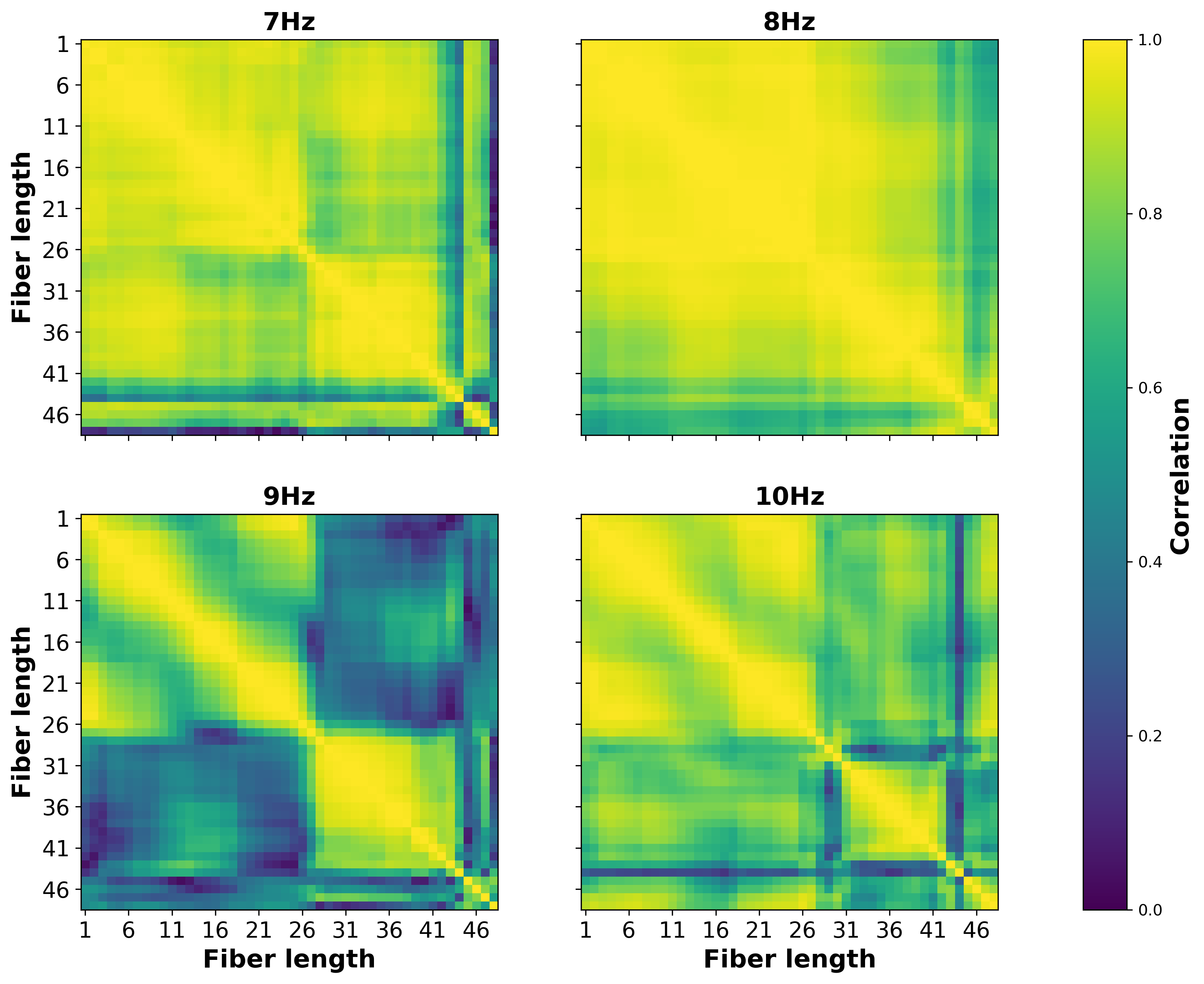}
	\end{minipage}
	\hfill
	\begin{minipage}{0.49\textwidth}
		\centering
		\includegraphics[width=\linewidth]{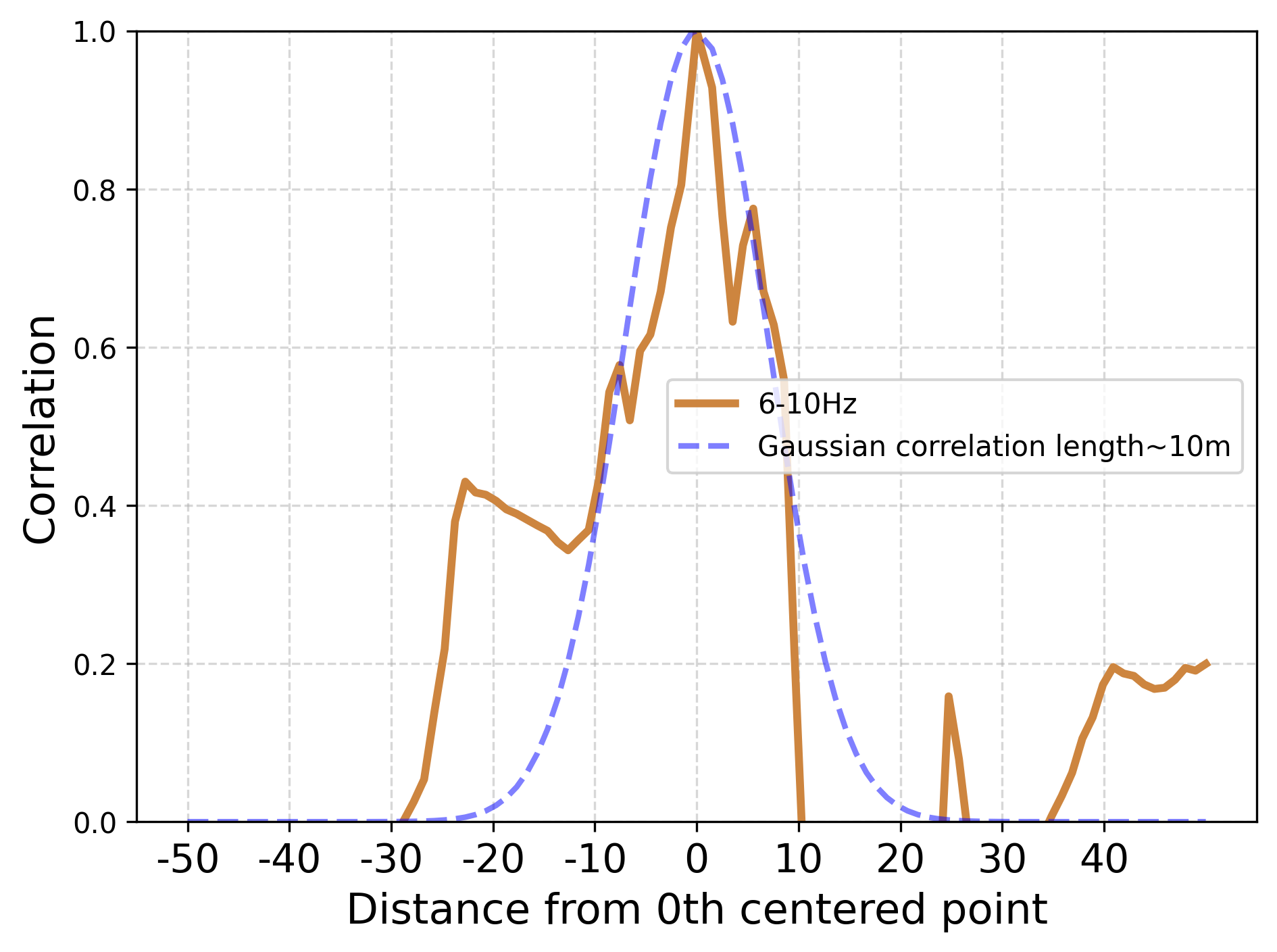}
	\end{minipage}
	\caption{\textit{Coherence length analysis of DAS in the 7–10 Hz band. Left: Coherence plots at 7, 8, 9, and 10 Hz indicating a coherence length of approximately 23 meters. Right: Comparison between the DAS correlation and the Gaussian correlation model, with the Gaussian correlation length estimated at 23.55 meters, closely matching the DAS-derived value}}
	\label{fig399}
\end{figure}

In the frequency range of 6–10 Hz, the coherence length of DAS increases to approximately 23 meters, as shown in figure \ref{fig399}. Again, this is consistent with the Gaussian correlation length, which is determined to be 23.55 meters, confirming the reliability of DAS for measuring spatial coherence at higher frequencies.

\vspace{0.2cm}

\noindent These findings suggest that DAS provides spatial coherence measurements with a resolution sufficient to characterize ground motion at scales smaller than 25 meters. Consequently, DAS offers an alternative or a compliment to traditional discrete seismic arrays for resolving local slab motion near gravitational wave detector optics. Its dense spatial sampling allows for improved modeling of ground motion without the need for densely deployed physical sensors, such as seismometers.

\section{Environmental Noise Monitoring with DAS}
One advantage of DAS is its ability to transform a standard optical fiber into a dense array of virtual sensors distributed along its entire length. Its high spatial resolution, wide bandwidth, and ability to cover long distances make it well-suited for detecting and characterizing both anthropogenic, natural acoustic, and seismic events. In this section, we demonstrate the capabilities of DAS in identifying various environmental noise sources—such as heating, ventilation, and air conditioning (HVAC) systems, and  thunderstorms using data collected from the DESY research campus.

\subsection{Detection of Mechanical Vibrations from HVAC Systems}
\begin{figure}[htbp]
	\centering\includegraphics[width=14cm]{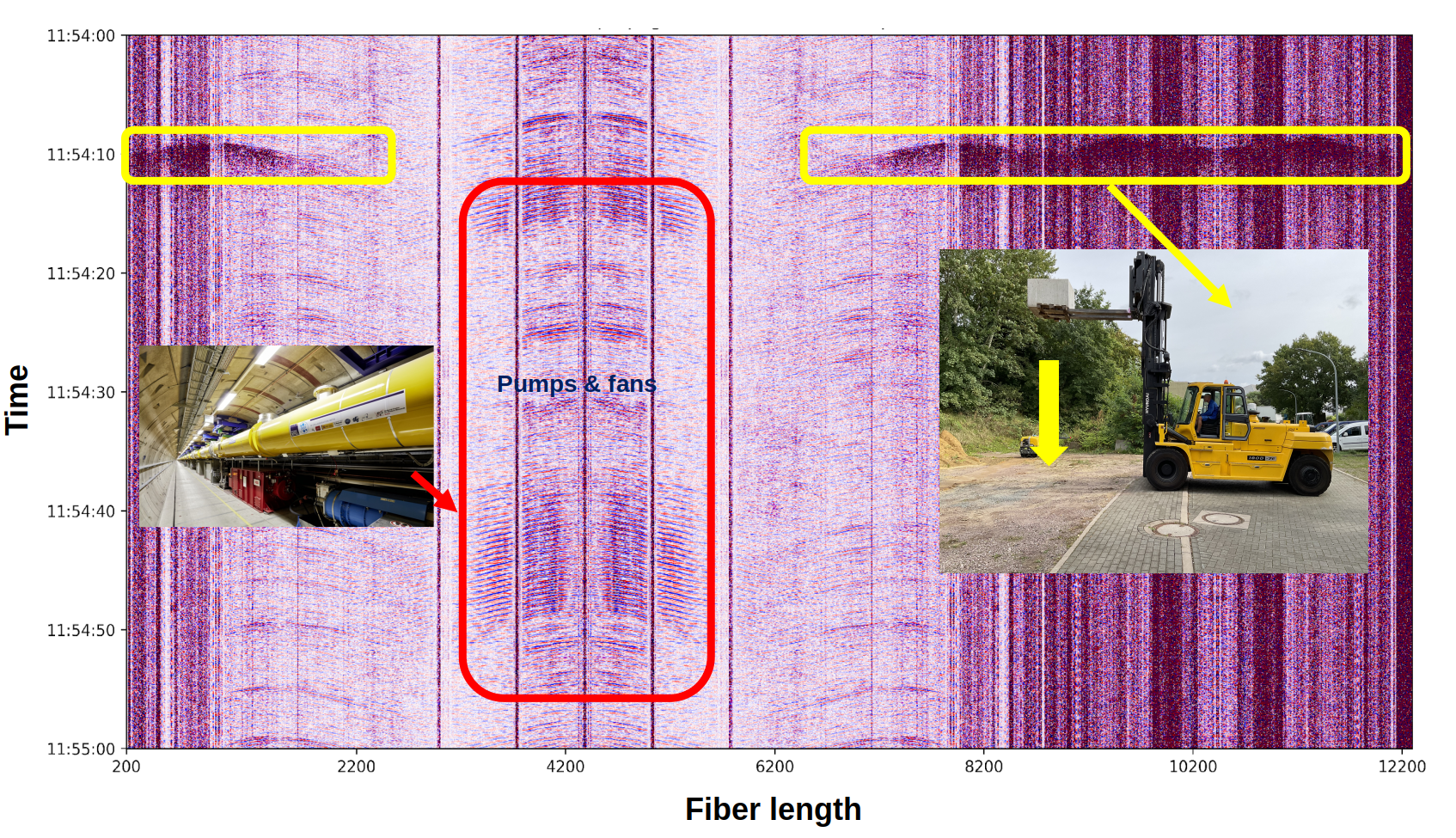}
	\caption{\textit{Illustration showing vibrations generated by HVAC systems (fans) within the European XFEL facility, as recorded by DAS. These persistent narrowband signals highlight DAS sensitivity to mechanical infrastructure. The right side displays DAS data capturing a controlled stone drop experiment. This illustrates the system’s ability to detect and localize low-frequency environmental noise with high spatial resolution.} }
	\label{fig12}
	
\end{figure}

DAS exhibits high sensitivity to low-frequency ground and structural vibrations generated by mechanical systems, such as  HVAC units. These systems often introduce narrowband vibration signatures into the surrounding soil or building structures. In our deployment, a dark fiber was routed inside the European XFEL, positioned along a rack running through the tunnel. This setup enabled the detection of HVAC machinery activation and deactivation as shown in figure \ref{fig12}.

\subsection{Detection of Thunderstorms and Atmospheric Acoustic Events}
\begin{figure}[H]
	\centering
	\begin{minipage}{0.6\textwidth}
		\centering
		\includegraphics[width=1.45\linewidth]{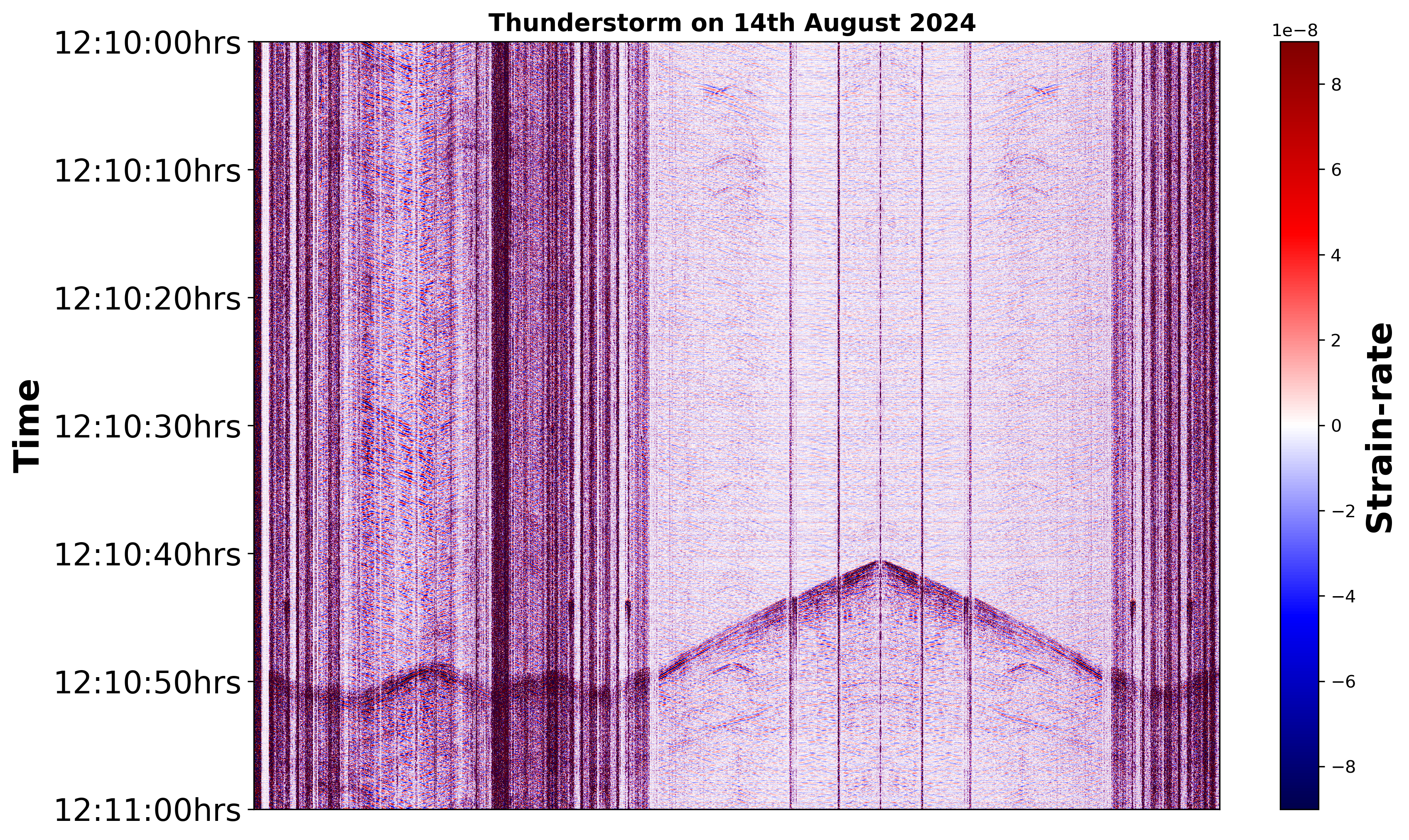}
	\end{minipage}
	\hfill
	\begin{minipage}{0.6\textwidth}
		\centering
		\includegraphics[width=1.45\linewidth]{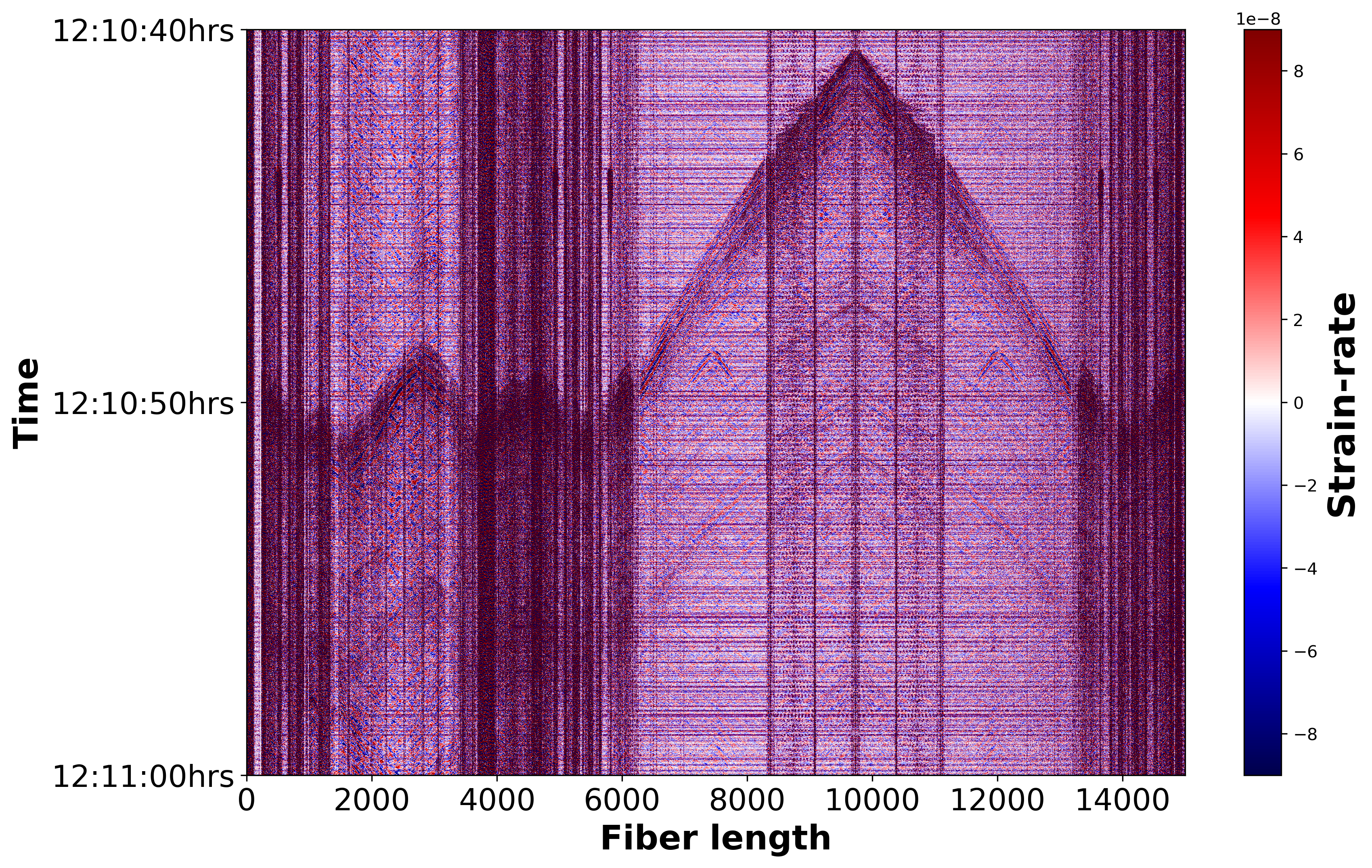}
		
	\end{minipage}
	\caption{\textit{Top: Illustration demonstrating the capability of DAS to detect atmospheric acoustic events.  Between sections 6000 and 10000, the wave arrival times are earlier compared to sections 500 to 6000. This time difference is attributed to faster wave propagation along the segment of fiber installed on the rack and ground pavement inside the tunnel. Below: Enlarged view revealing the decaying coda of the thunderstorm signal, illustrating the ability of DAS showing gradual attenuation of energy over time.}}
	\label{fig13}
\end{figure}

DAS also demonstrated the ability to capture transient, broadband signals generated by atmospheric disturbances such as thunderstorms. During a storm event recorded at the DESY site, the DAS array detected impulsive signals across multiple fiber channels that corresponded in time with lightning strikes and thunderclaps confirmed by external weather data as shown in figure \ref{fig13}.

\vspace{0.2cm}

\noindent These signals were characterized by high amplitude, high-frequency components, followed by decaying coda waves. The spatial footprint of these events was extended, often affecting tens to hundreds of meters of the fiber array, allowing the reconstruction of wavefront arrival angles and propagation speeds. Such data can be used not only for event detection but also for atmospheric monitoring and characterizing ground-atmosphere coupling.

\section{Conclusion}
\noindent In this study, we have demonstrated that DAS is a highly capable tool for seismic and environmental monitoring in the context of gravitational wave detection. Our results show that DAS can not only serve as an alternative to traditional seismic sensors such as geophones and broadband seismometers but also act as a complementary system that enhances spatial coverage and overall sensitivity. We validated the fidelity of DAS by converting its measurements into seismometer-equivalent signals, confirmed through PSD comparisons, Bland–Altman analyses, and additional statistical evaluations.

\vspace{0.2cm}

\noindent By applying multichannel Wiener filtering, we successfully predicted geophone responses from DAS data, demonstrating the accuracy and coherence necessary for Newtonian noise cancellation strategies. As a case study, we showed that DAS can effectively cancel noise recorded by the vertical component of a seismometer, achieving a residual factor of 0.11 at 20 Hz—highlighting its potential role in precision noise mitigation.

\vspace{0.2cm}
\noindent Beyond sensor replication, DAS exhibits very good sensitivity to a wide range of environmental and anthropogenic noise sources, including microseisms, wind events, thunderstorms, and passing vehicles. This broad detection capability is important for gravitational wave observatories, where real-time noise characterization and mitigation are essential for enhancing low-frequency sensitivity.

\vspace{0.2cm}

\noindent Building on insights gained during this work—including discussions on coherence, filtering strategies, and sensor optimization—our findings suggest that DAS, due to its scalability, use of existing fiber infrastructure, and dense spatial sampling, is well-positioned to support and enhance seismic and environmental monitoring for current and future detectors like LIGO and the Einstein Telescope.

\section{Acknowledgment}
\noindent  These experiments were
conducted in collaboration with the WAVE Initiative Hamburg,
a joint effort involving Helmut Schmidt University, University of Hamburg, Deutches Elektronen- Synchrotron,
GeoForschungszentrum Potsdam, and the European XFEL. The authors would like to acknowledge all the individuals
involved in the WAVE initiative.

\vspace{0.2cm}

\noindent Special thanks goes to Prof. Jan Harms for helping to understand Newtonian noise cancellation.

\section{Appendix}\label{intro}
\noindent
To derive Eq.~(\ref{eq8}), we begin with the following assumptions and definitions.

\vspace{0.2cm}

\noindent Consider an array of M sensors sensors distributed along an optical fiber of total length L. Let the positions of the sensors be denoted by $D_{1}$,$D_{2}$,$D_{3}$,$\cdots$,$D_{M}$, where $D_{1}$ corresponds to the first sensor and $D_{M}$ to the last. The distance between any two sensors in any direction, for example between the first and second, is given by $\left| D_{2}-D_{1} \right|$.

\vspace{0.2cm}

\noindent Suppose a seismic wavefront $K$ impinges on the array. It arrives at the first sensor at time $t_{1}$, at the second at time $t_{2}$ and at the $i$-th sensor at time $t_{i}$. For the $M$-th sensor, the arrival time can be expressed as:
\begin{equation}
	t_{M}=\frac{D_{M}-D_{1}}{V_{p}}, 
\end{equation} where $V_{p}$ is the seismic particle velocity and its inverse is the seismic slowness  $s$.

\vspace{0.2cm}

\noindent
This time-delay relationship allows us to align the seismic signals across the sensor array by shifting each channel's time series based on a tested slowness value. When the assumed slowness matches the true slowness of the wavefront, the aligned signals will constructively interfere, producing a peak in the semblance value.

\noindent
The semblance coefficient is then computed as:
\begin{equation}
	\text{Semblance}(s, t) = \frac{1}{M} \cdot \frac{\left[ \sum_{i=1}^{M} x_i(t + \Delta t_i(s)) \right]^2}{\sum_{i=1}^{M} x_i^2(t + \Delta t_i(s))},
	\label{eq16}
\end{equation} where $x_{i}(t)$
 is the time series from the 
$i$-th sensor, and 
$\Delta t_{i} (s)$ is the time delay needed to align the 
$i$-th sensor’s signal with the reference sensor based on the tested slowness $s$. A high semblance value indicates high coherence among the aligned signals, which in turn implies a consistent wavefront propagating across the array at the assumed slowness.

\vspace{0.1cm}

\noindent
Following the semblance definition from \cite{b8a}, the semblance at index $j$ is given by:
\begin{equation}
sem[j]=\frac{[\sum_{i=1}^{M} f(i,j)]^{2}+ [\sum_{i=1}^{M} h(i,j)]^{2}}{M \sum_{i=1}^{M} \{ f(i,j)^{2} + h(i,j)^{2}\} },
	\label{eq78}
	\end{equation} where $f(i,j)$ and  $h(i,j)$ represent the real and imaginary components of the signal, respectively.

\vspace{0.1cm}

\noindent Substituting Eq.(\ref{eq16}) into Eq.(\ref{eq78}) and  through straightforward algebraic manipulation, leads to Eq.~(\ref{eq8}).

\appendix


\begin{thebibliography}{00}
	\bibitem{b1} A. H. Hartog, “An Introduction to Distributed Optical Fibre Sensors". CRC Press,
	2017.
	
	\bibitem{b2a} J. Harms, “Terrestrial gravity fluctuations”, Living Reviews in Relativity, 2019.
	
	\bibitem{b2b}R. Rading, K.S. Isleif, “Coherent and Incoherent Noise Cancellation using Distributed Optical Fiber Sensors", IEEE
	Photonic Society Summer Topicals, 2024.
	
	\bibitem{b3} Y. Li, M. Karrenbach, and J. B. Ajo-Franklin, “A Literature Review:
	Distributed Acoustic Sensing (DAS) Geophysical Applications Over the Past 20 Years”, pg. 229–291, 2021.
	
	
	\bibitem{b4} S. Hughes and K. Thorne, “Seismic gravity-gradient noise in interferometric gravitational-wave detectors”, Physical Review D 58.12, 
	1998.
	
	\bibitem{b15a} A.H. Hartog, “Propagation in optical fibres,” in An Introduction to Distributed
	Optical Fibre Sensors, CRC Press, 2017.
	
	\bibitem{b16a} S.D. Personick, “Photon probe — an optical-fiber time-domain reflectometer,”
	The Bell System Technical Journal, vol. 56, no. 3, pp. 355–366, 1977.
	
	\bibitem{b14a} SEAFOM. “DAS parameter definitions and tests.”, [Online]. Available:
	http://seafom.com.
	
	\bibitem{b9a}H.F. Wang,  X. Zeng,  D.E. Miller, D. Fratta,  K.L. Feigl,
	C.H. Thurber, and R.J. Mellors,“Ground motion response to
	an ML 4.3 earthquake using co-located distributed acoustic sensing and seismometer arrays", Geophys. J. Int., 213, 2020–2036, 2018.
	
	\bibitem{b10a}  T.M. Daley, D.E. Miller, K. Dodds, P. Cook, and  B.M. Freifeld, “Field
	testing of modular borehole monitoring with simultaneous distributed acoustic sensing and geophone vertical seismic profiles at Citronelle",
	Alabama, Geophys. Prospect., 64(5), 1318–1334, 2015.
	
	\bibitem{b11a}  N.J. Lindsey,  H. Rademacher, and  J.B. Ajo-Franklin, “On
	the Broadband Instrument Response of Fiber-Optic DAS"
	Arrays, J. Geophys. Res.-Sol. Ea., 125, 2020.
	
	\bibitem{b11b} L.N.B. Kennett. “A Guide to the Seismic Wavefield as seen by DAS", Research School of Earth Sciences, Australian National University, 2024.
	
	

	
	\bibitem{b6a} M.T. Taner,  F. Koehler, and  R.E. Sheriff, “Complex
	seismic trace analysis", Geophysics, 44, 1041–1063, 1979.
	
	\bibitem{b7a}  N.S. Neidell and  M.T. Taner, “Semblance and other Coherency
	Measures for Multichannel Data", Geophysics, 36, 482–497, 1971.
	\bibitem{b8a}  T. Shi and  S. Huo,“Complex Semblance and Its Application", J. Earth Sci., 30, 849–852, 2019.
	
	\bibitem{b12a}  R. Schmidt, “Multiple emitter location and signal parameter estimation", IEEE Trans. Antennas Propag., 34(3), 276–280, 1986.
	
	\bibitem{b13a}  M. Hobiger,  C. Cornou,  P.-Y. Bard,  N. Le Bihan, and W. Imperatori,
	“Analysis of seismic waves crossing the Santa Clara valley using the three component MUSIQUE array algorithm", Geophys. J. Int., 207(1), 439–456, 2016.
	
	\bibitem{bb2} WAVE Initiative: https://wave-hamburg.eu/
	
	
	\bibitem{b14} F. Badaracco, “Newtonian Noise Studies in 2nd and 3rd Generation Gravitational Wave Interferometric Detectors", PhD thesis, Gran Sasso Science Institute, 2020.
	
	\bibitem{b15}A. Basalaev, “Spicypy”. Zenodo, doi: 10.5281/zenodo.13947023, 2024.
	
	

	
	
	

\end{thebibliography}
\end{document}